# Unified Control and Data Flow Diagrams Applied to Software Engineering and other Systems

by Igor Polkovnikov
2016, October 3
San Francisco, CA, USA

## Introduction

More often than not, there is a need to understand the structure of complex computer code: what functions and in what order they are called, how information travels around static, input, and output variables, what depends on what. As a rule, executable code and data are scattered among multiple files and even multiple modules. Information is transmitted among variables which often change names. These tangled relations greatly complicate the development, maintenance, and redevelopment of code, its analysis for complexity and its robustness. As of now, there is no tool which is capable of presenting the real-life, useful diagram of actual code. Conventional flowcharts fail. Proposed is the method which overcomes these difficulties. The main idea is that functionality of software can be described through flows of control, which is essentially flows of time, and flows of data. These are inseparable. The second idea is to follow very strict system boundaries and distinctions with respect to modules, functions, blocks, and operators, as well as data holders, showing them all as subsystems, in other words, by clearly expressing the system structure when every piece of executable code and every variable may have its own graphical representation. The third is defining timelines as the entities clearly separated from the connected blocks of code. Timelines allow presentation of nesting of the control flow as deep as necessary. As a proof of concept, the same methods successfully describe production systems.

If one intends to make a diagram which reflects the structure of the current code, the data and control flows must both be present. Since these flows are fundamentally different, they are represented by two different directed lines:

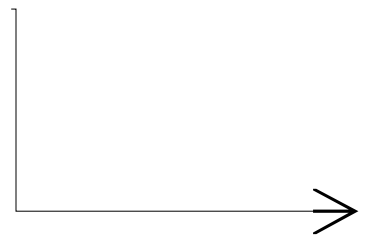

- A straight angled thin line represents control flow.

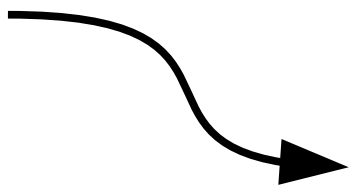

- A curved thicker line represents data flow.

Thinking philosophically, control flow is a flow of the time allotted for different components. Most software processes have essentially a sequential nature, but, of course, there are “parallel” ones. Hardware works in such a way as to grant time from higher-level processes, or functions and operators, to the lower-level ones. If so, it is then possible to say that “time flows from one component to another”.

Data flow is a flow of information from one data holder to another. A data holder is a register, memory location, or even a signal and therefore a variable, file, paper, and so on. We say that information flows from one location to another. Mechanisms providing this are processes.

Both data and control flows are essential to understanding the structure and the way of operating of a software process, on both a small and on a large scale. Representing it with a diagram was the dream of many generations of software developers, especially those who “think visually”. The complexity of even simple pieces of code prevented it from becoming successful. Many attempts have been made to diagram the software, but it has ether drowned in unnecessary details or is too simple and does not represent the real system with all its bugs, unintended logical and hardware dependencies, and language particularities. It is therefore, of little value in solving practical problems. Many attempts has been made to overcome these difficulties. Let me present one more.

It is hard to believe that a diagram is capable of representing the entire complexity of the relations of any working software. If it is made, because of its vastness, humans would  not be capable of comprehending it in detail, unless it is of interest to overview the system as whole, grasping the intuitive properties that are most visible. That can be useful. However, in practice, a hierarchical way would be preferred, where the diagram complexity is chosen based on a task to solve. Therefore the same diagramming method has to be able to present the lower level, down to single operators or even the assembler, up to functions, classes, and processes within a single program, computer, or network. Abbreviations have to be identical. If so, it means that the abstraction of underlying information processes have been found. If the same method could be applied to other systems such as devices, machinery, tools, and production lines, that would support the strength and correctness of the method.

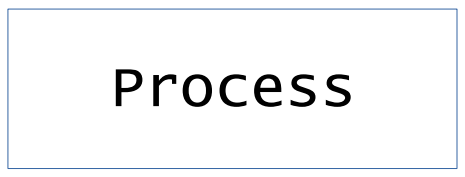

- A rectangle represents a process. (A “process” in the most general meaning)

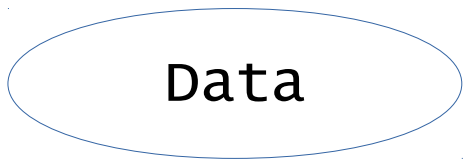

- An ellipse represents a data holder. Software specifies data holders as physical entities. Information is transferred among data holders by means of processes. Note, that method presented here does not deal with the structure of information. It deals with physical structure of data holders. The informational content of a data holder can be represented in several ways, see later.

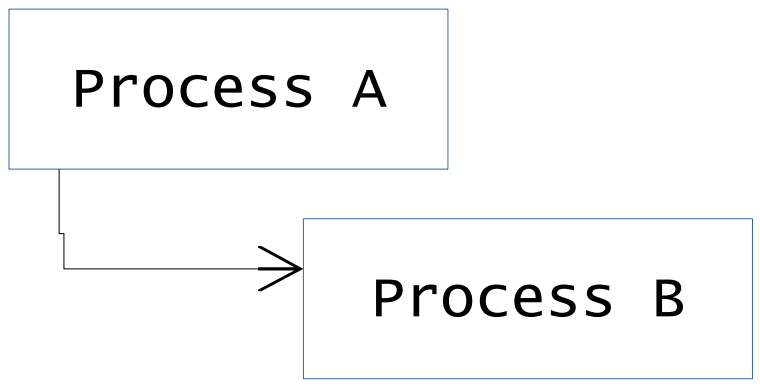

- Conrol may flow only among processes. The figure means that B is a subprocess of A.

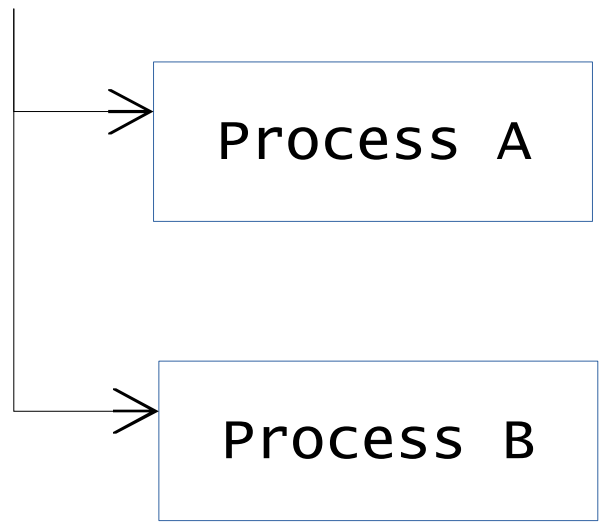

- What is shown here means processes A and B are parallel.

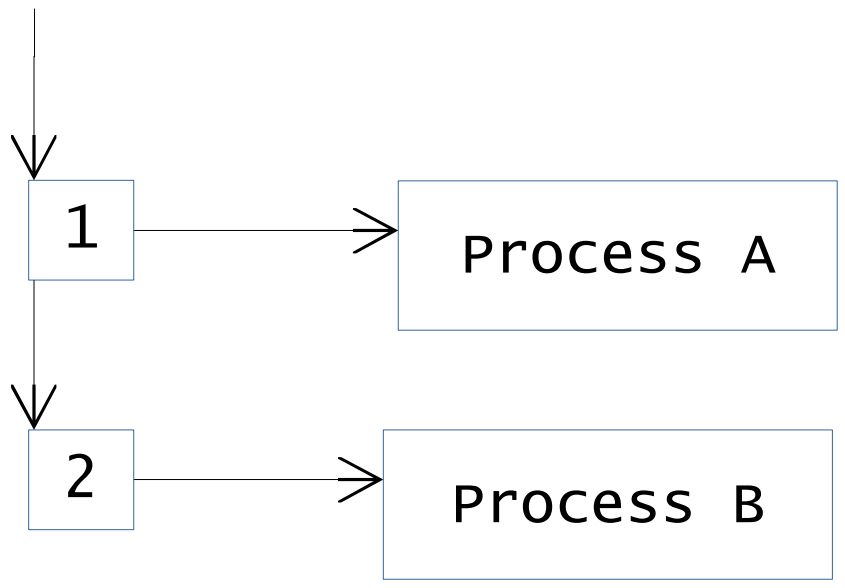

To specify the order of processes, a control flow line should be marked with numbers or anything that specifies the order in small rectangles, for example, like this.

- This means, first process A, then B. The line of order may be called a “timeline”. The major direction of the timeline line is shifted to distinguish it with a line directed toward the next process block.

There is a system behind any timeline serving its operation. In computer, it is a processor.

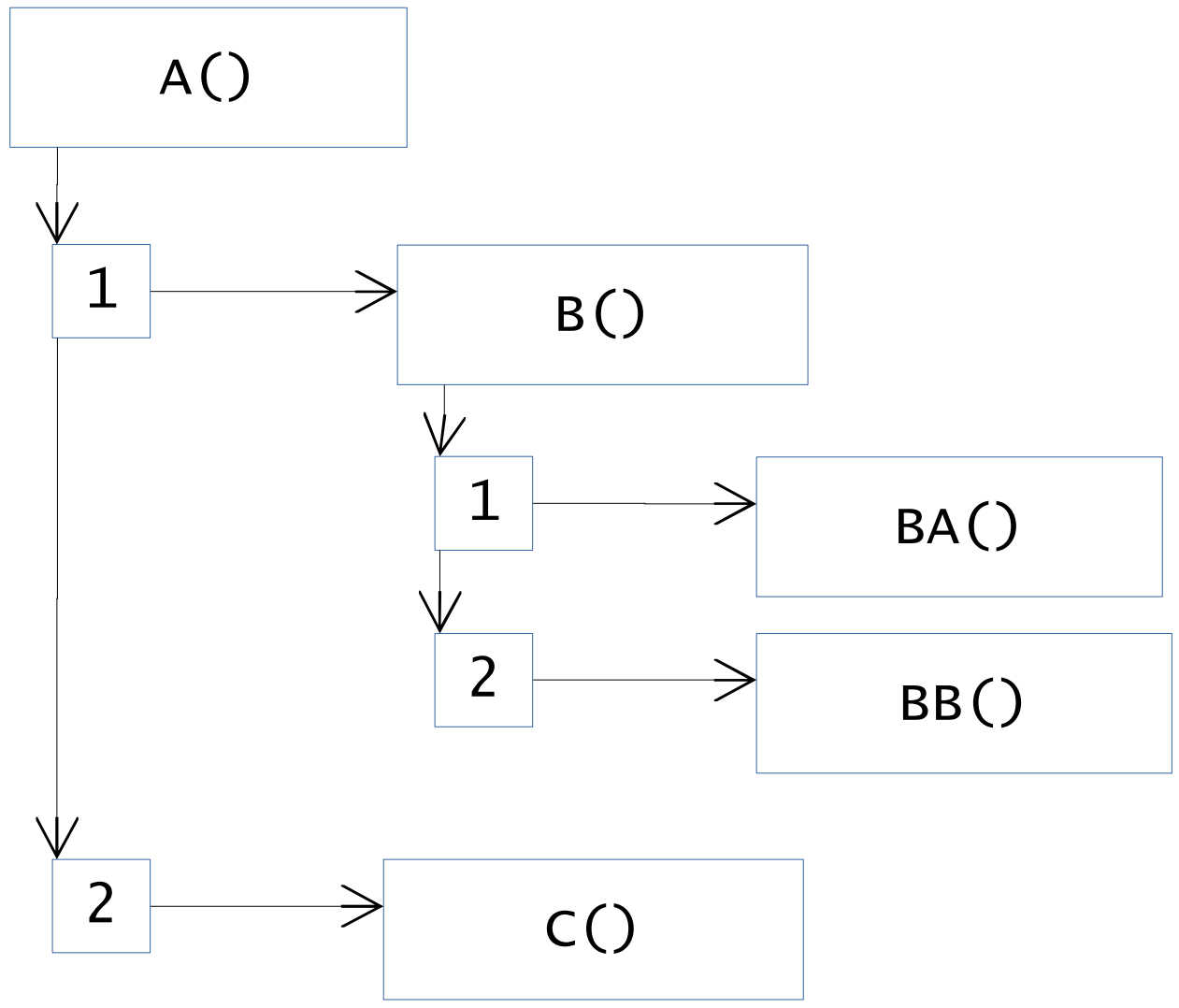

The timeline represents an underlying mechanism of time distribution. Timelines allow for a clear separation of, for example, nested function calls.

***Timelines are one of the major differences with common flowcharts or control diagrams. They allow the showing of the main process as well as as many nested functions or blocks as required.***

Process blocks connected by the same timeline belong to the same system level. A new timeline coming out of a block (out of B(), on this example) describes a subprocess (a subsystem level), etc. as deeply nested as required.

Timeline numbering may vary. A time stamp serves well when required.

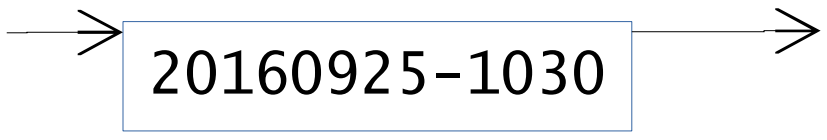

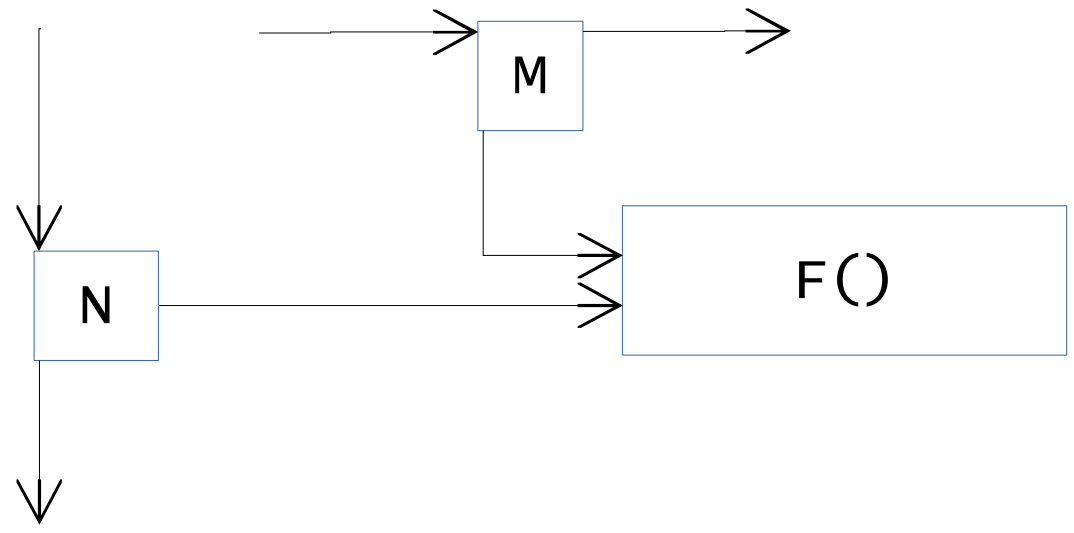

Timelines can show how a single function is called from different places. Here M and N specify an order on two different timelines from which F() is called.

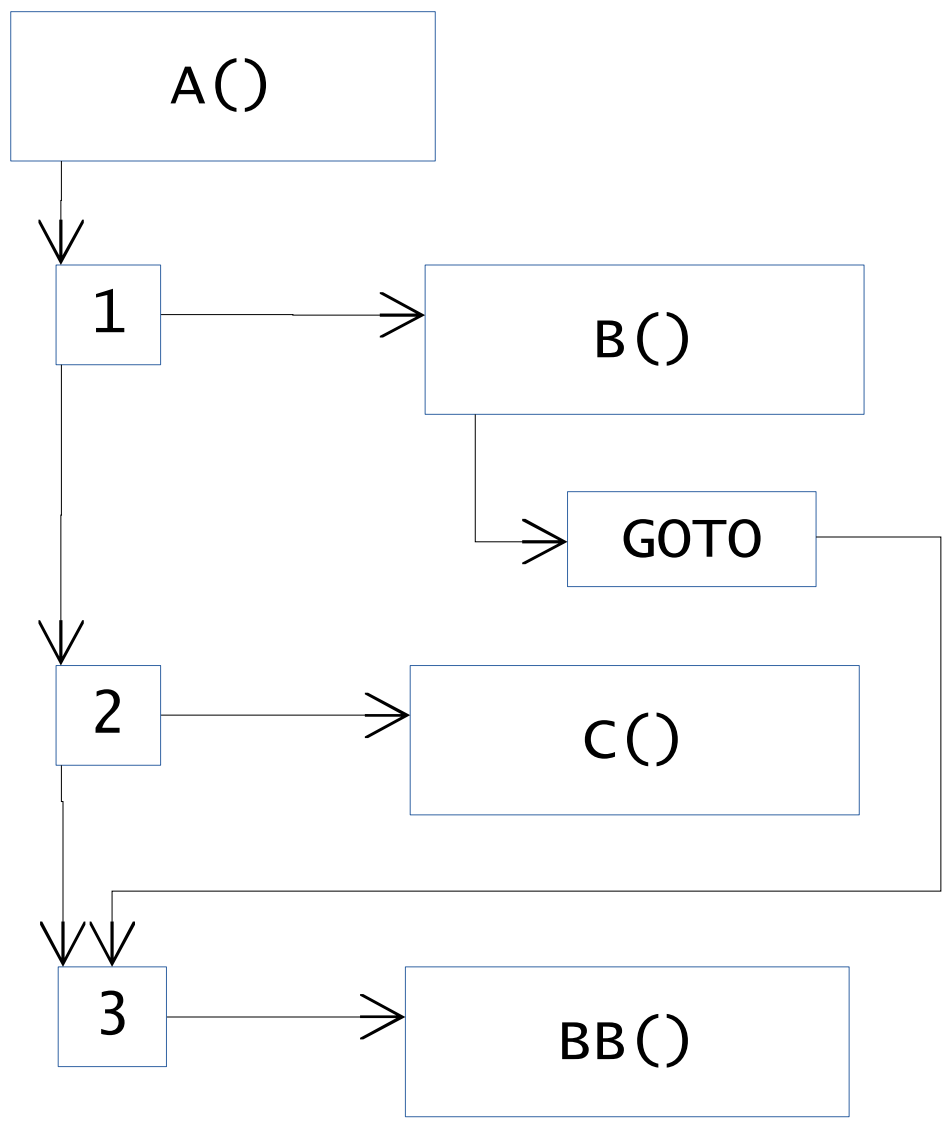

Regarding most computer languages, if a block has the GOTO statement (or any kind of it), there is no difficulty in presenting its meaning with timelines on a drawing.

GOTO type control statements (other examples are “continue”, “break”, and so on) and control flow lines coming out of them are different from the regular process blocks depicting functions, {}-code blocks, and operators. Those “return” to the timeline rectangles. The control line, leading to the “returning” process blocks, should be double-arrowed. It may be omitted for simplicity as it is done in this example. However if there are too many GOTOs, there is no other way to avoid confusion, but to use double-arrowed lines.

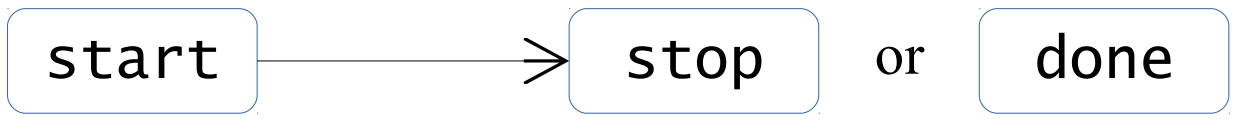

Timelines can be marked with the ‘start” and/or “stop”, or “done” marks for clarity.

The purpose of the process blocks is to update, generate, convey information, create and delete data holders as well as to modify the structure of the timelines. Here is how it is shown.

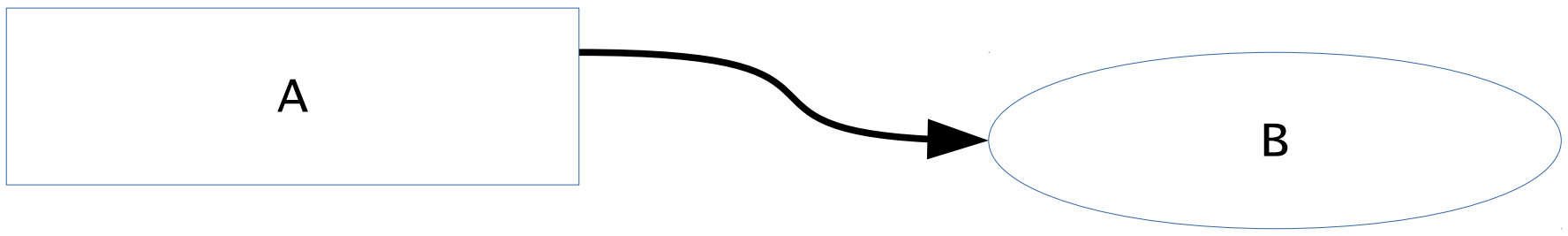

- This means the process A updates information in the data holder B.

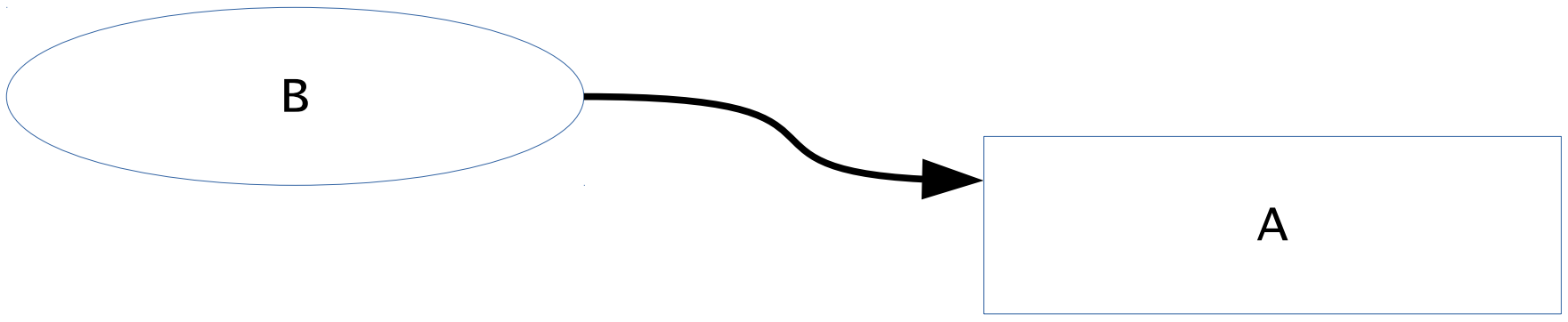

- This means the process A is using information from the data holder B.

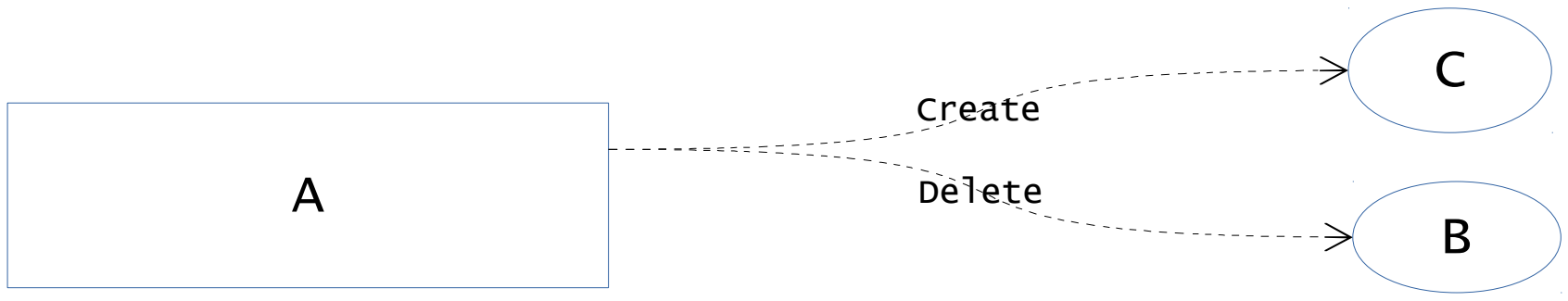

- This means the process A creates the data holder C and destroys the data holder B.

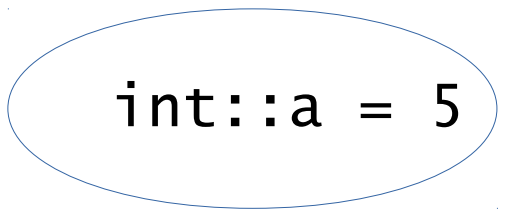

- Data holders often have names. The names and the actual information content can be shown simply. A name may be complex and include the type name, for example, “int::”.

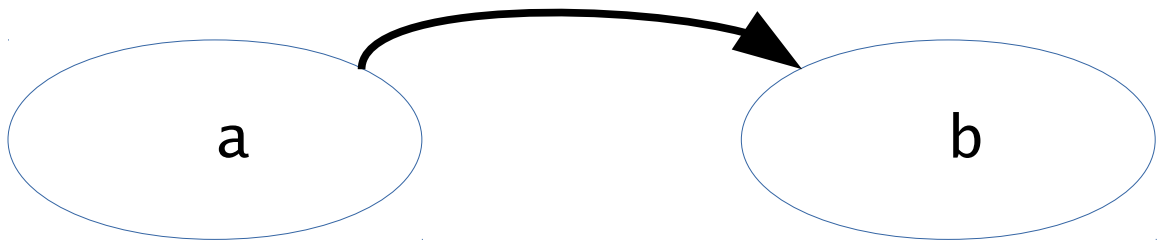

- If there is a flow of information between data holders without regard to how it is done, it is presented like this. Likewise, a data flow arrow among rectangles represents a data flow among processes without regard to the details: it exists.

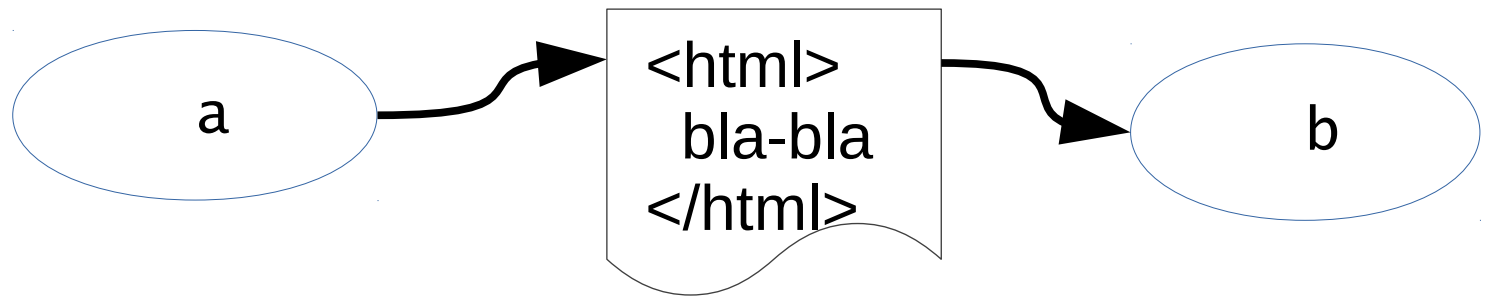

- Sometimes there is a need to show actual content of information without regard to its holders. It can be done with a document shape (see later), since by writing an information content with symbols (any symbols), we are creating a document.

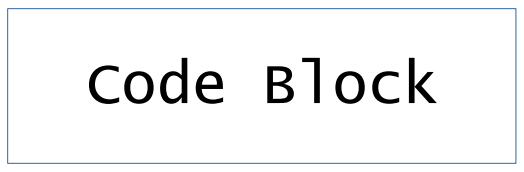

In software, processes may be different in methods of their implementation and action. Firstly, there are functions, blocks, and operators, with their sequences and combinations. Their calling order is precisely defined by code. They are shown with a simple rectangle.

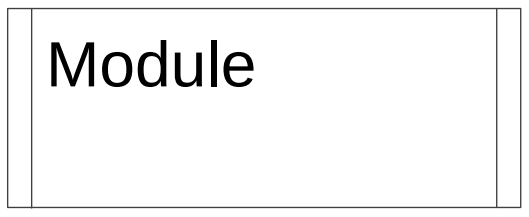

Secondly, there are object modules, files, classes, and so on, which may contain a variety of functions and variables that may be accessed independently or serve as templates to be instantiated. These processes, or rather process collections, are called “Modules”. On a broader view they are more complex systems. Let us represent them as rectangles with double-struck vertical edges.

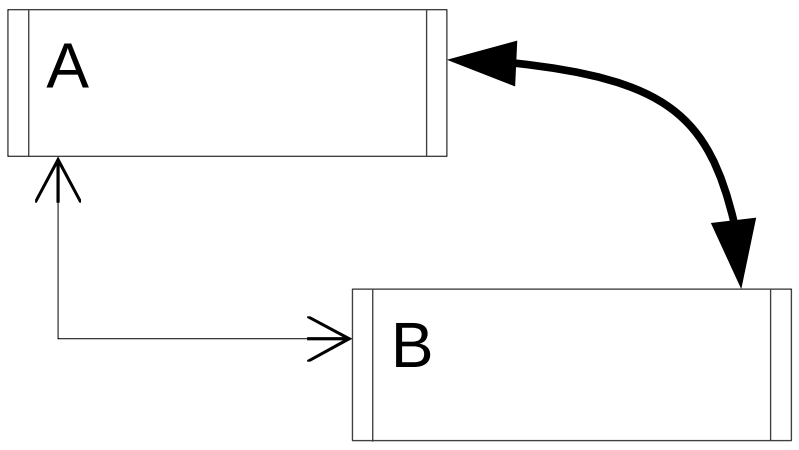

- When there is need to show interaction among modules or processes, simple arrows of both kinds can be used. The diagram on the left shows that internal code blocks of modules A and B are calling each other and they are exchanging information.

A timeline drawn to a module means that one of its “processes” is called. A data flow line to a module means one of the module’s variables has changed, or a module’s function receives an argument.

Different kinds of data holders may be presented by their own kinds of shapes.

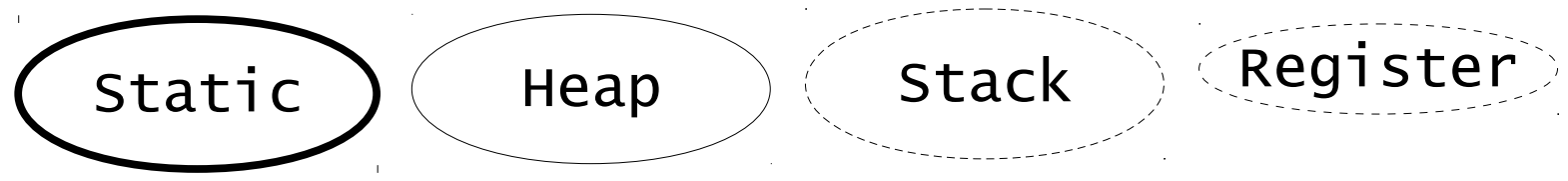

- Static, heap, stack, and register data holders.

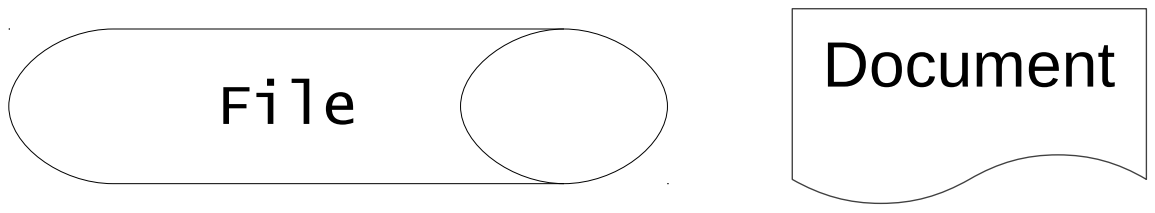

- A file and a document. File can be a file, a directory, or a drive. A document can be a printed one, a monitor, or something else.

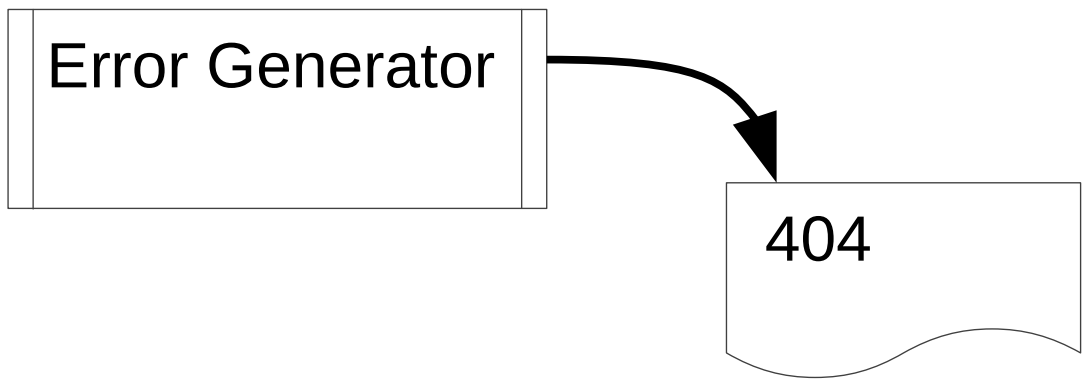

- For example, a module “Error Generator” updates a document called 404. A detailed mechanism of that is not shown.

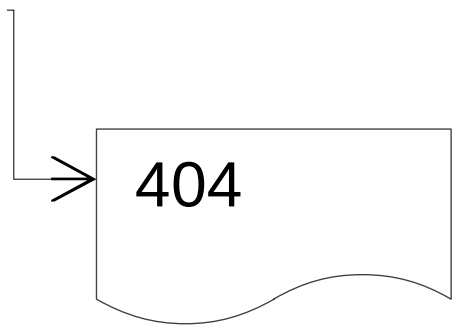

- Often, if it is needed to distinguish between human actions and automatic actions. A “paper” shape can be used to represent a human-driven action. It then depicts a process. I have found it convenient.

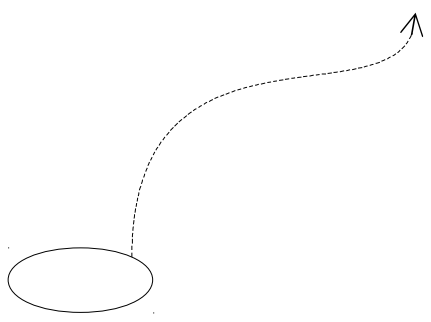

- If needed, pointers can be shown as small ellipses with a curved dotted arrow to a referenced entity. Pointers can be omitted, if they do not carry crucial information, like in arguments to function calls.

Both processes and data holders may have parts. A relation of “having a part” can be presented by another line.

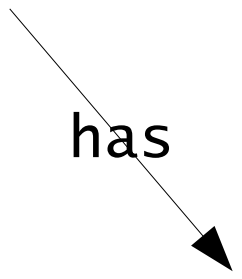

- it may be a straight angled or angled line to differentiate it from a control flow line.

Examples:

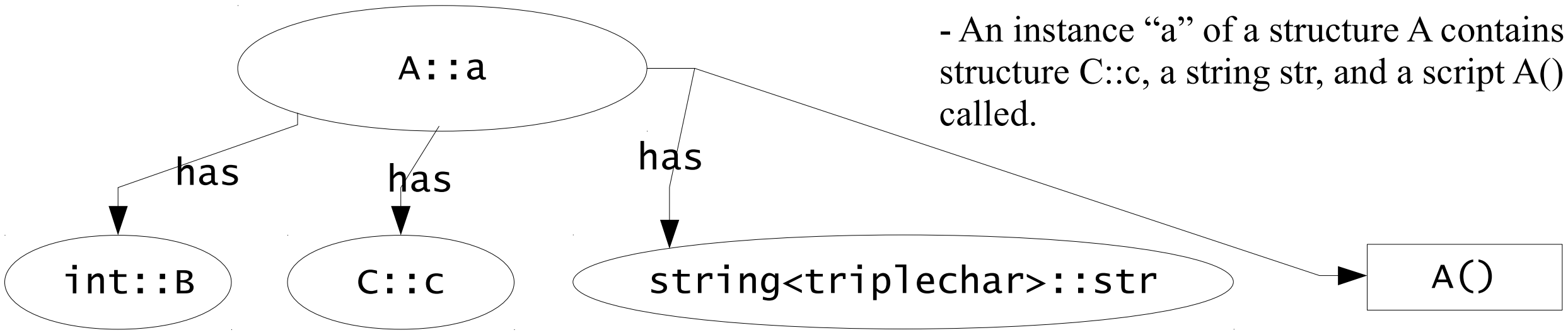

- An instance “a” of a structure A contains an integer B, a structure C::c, a string str, and a script A() which may be called.

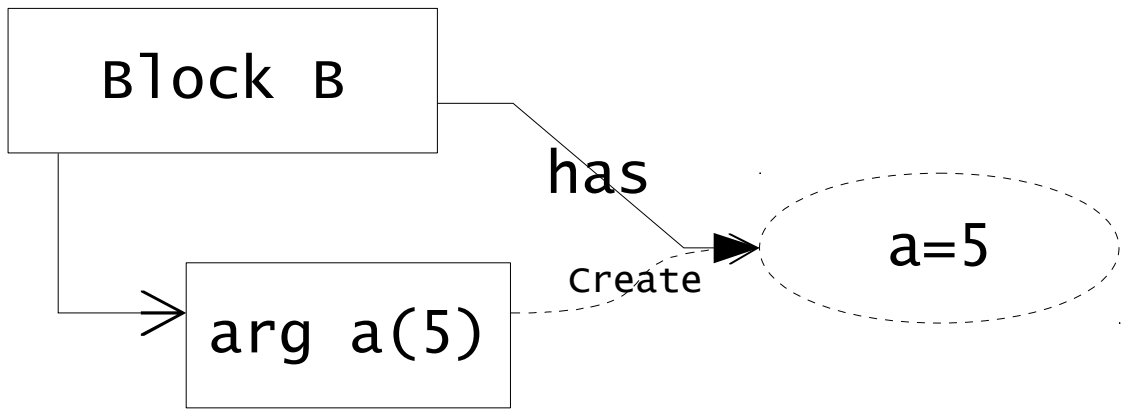

- A code block B has a variable “a” created on stack by a statement “arg a(5)”. The statement may be omitted:

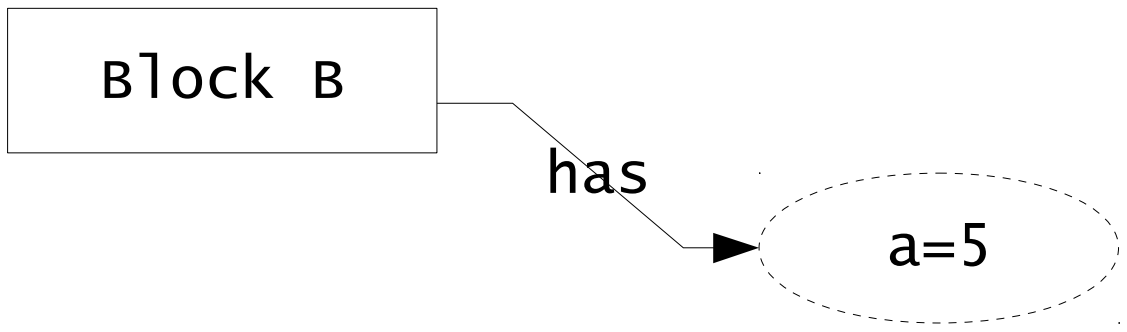

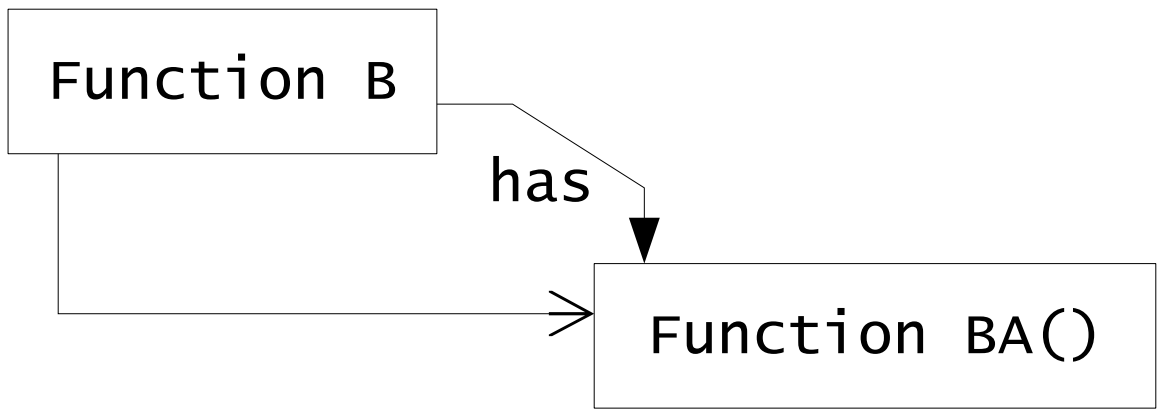

- A function B() defines a function BA() and calls it. It may be possible in MATLAB or JavaScript.

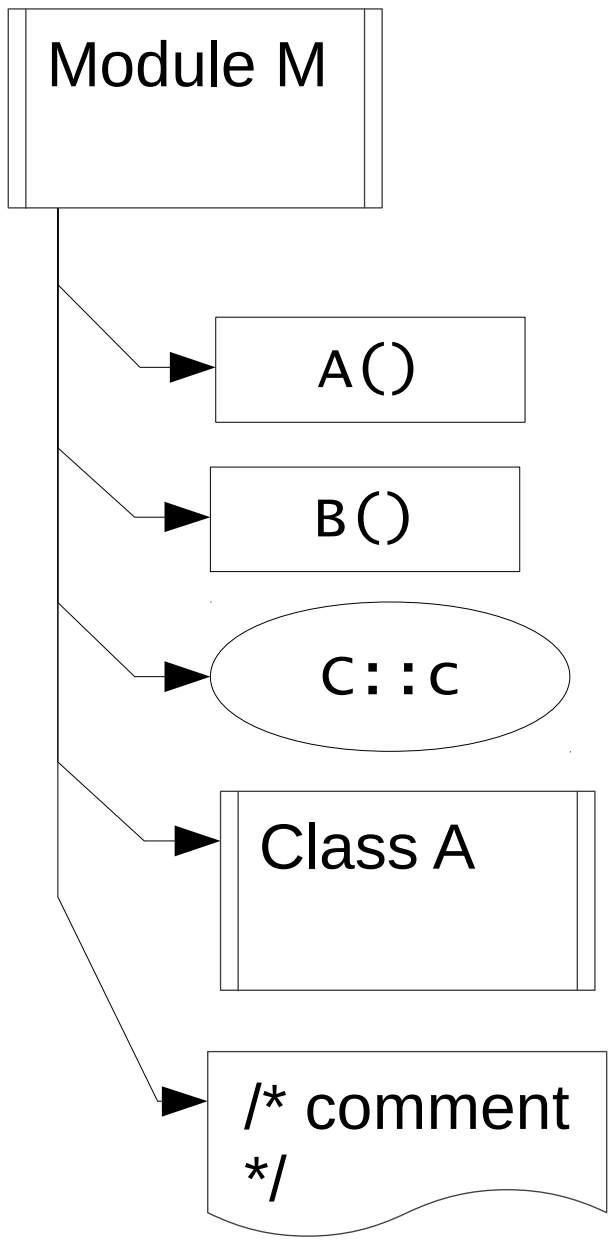

- A module M (may be a file compiling into an object module) contains functions A() and B(), an instance “c” of a class C, a class A, and a comment which might be a text in the source file or else.

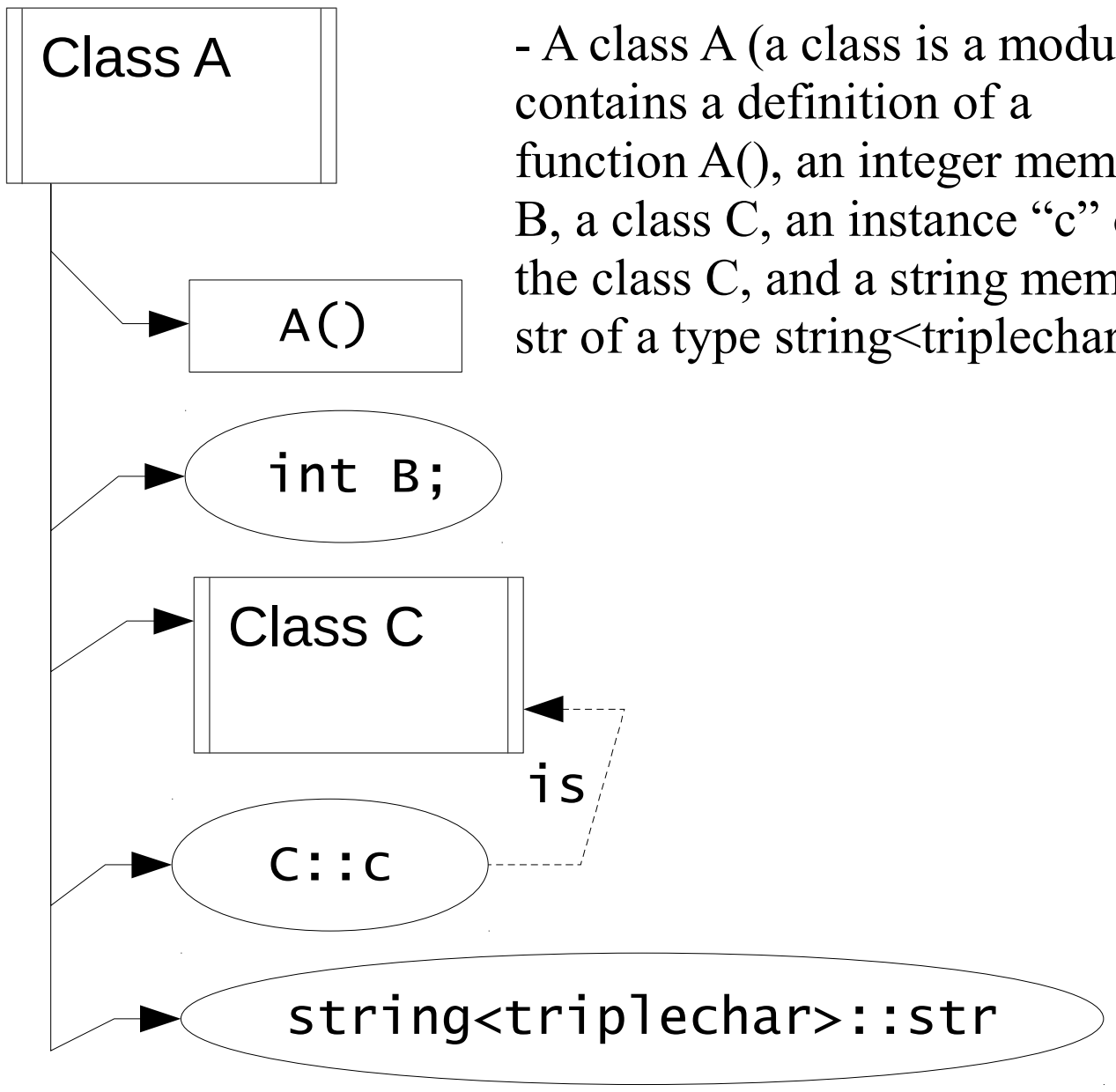

- A class A (a class is a module), contains a definition of a function A(), an integer member B, a class C, an instance “c” of the class C, and a string member str of a type string<triplechar>.

Sometimes the relation “has” is convenient to replace with a Euler diagram.

Examples:

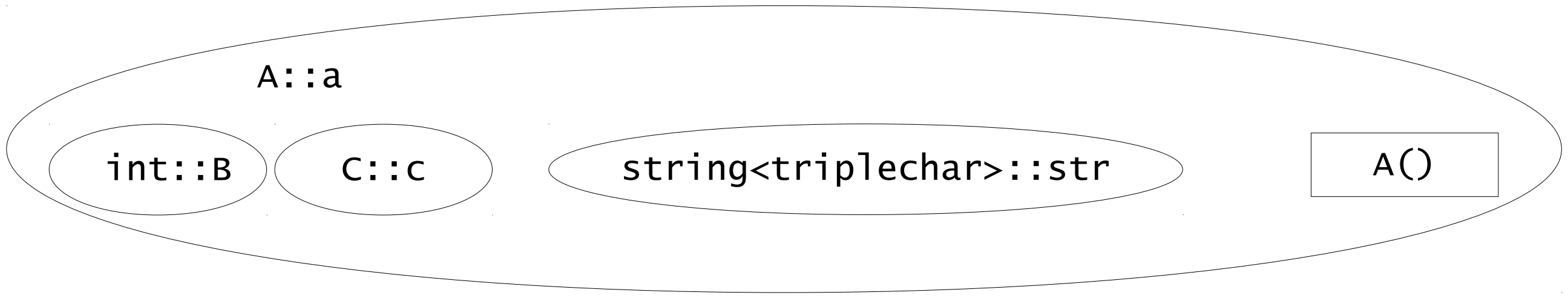

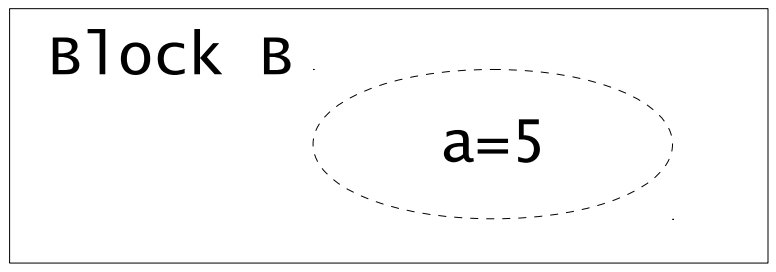

Both ways are useful. The Euler diagram method is more compact. The “has” line method allows to take out a part to a more convenient location, or quickly associate an existing data holder with a more complex one as being a part of it.

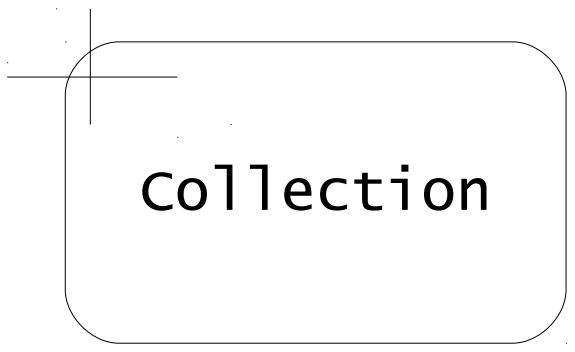

Collections implemented as, for example, arrays, lists, dictionaries, and so on, are data holders. They accept and store other data.

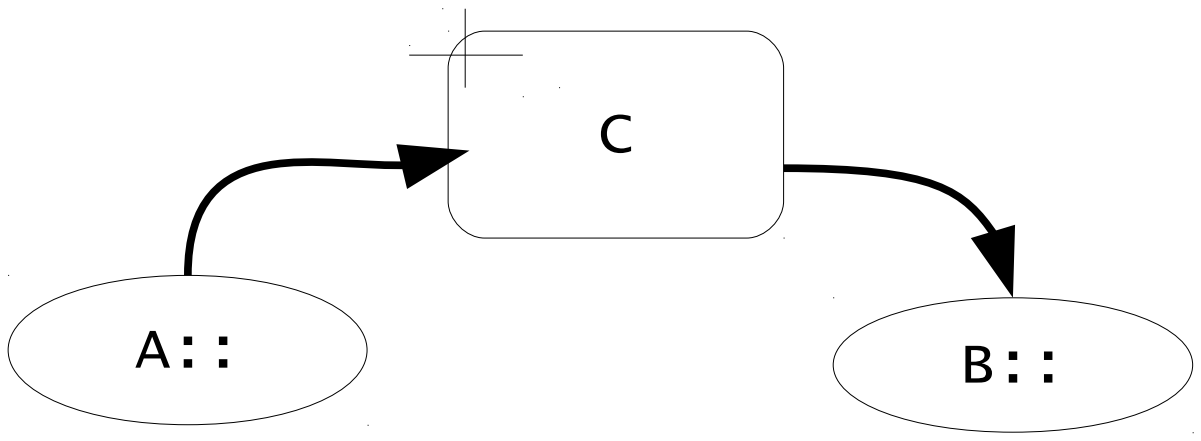

- For example, data of type A are stored in the collection C, data of type B are read from it.

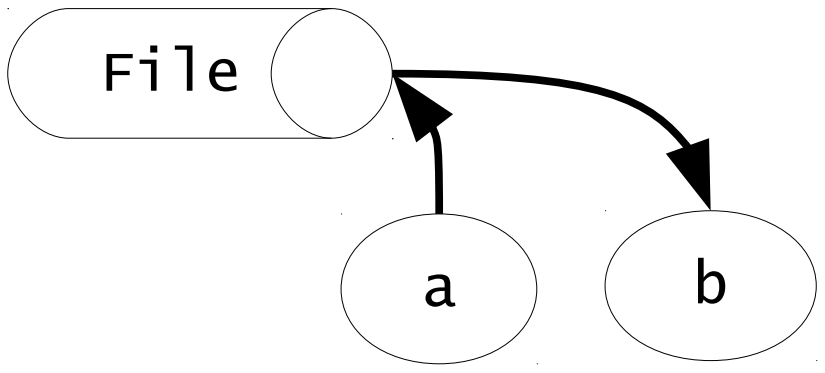

- Files are also collections.

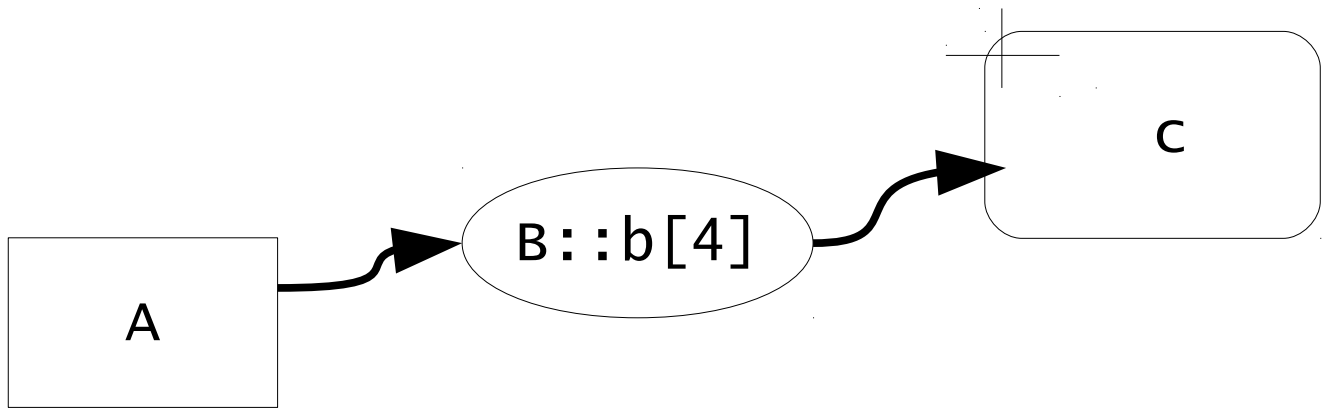

- Process A updates the record "b[4] of type B in the collection "c". The record b[4] may have "has" line to C or be a Euler part of C.

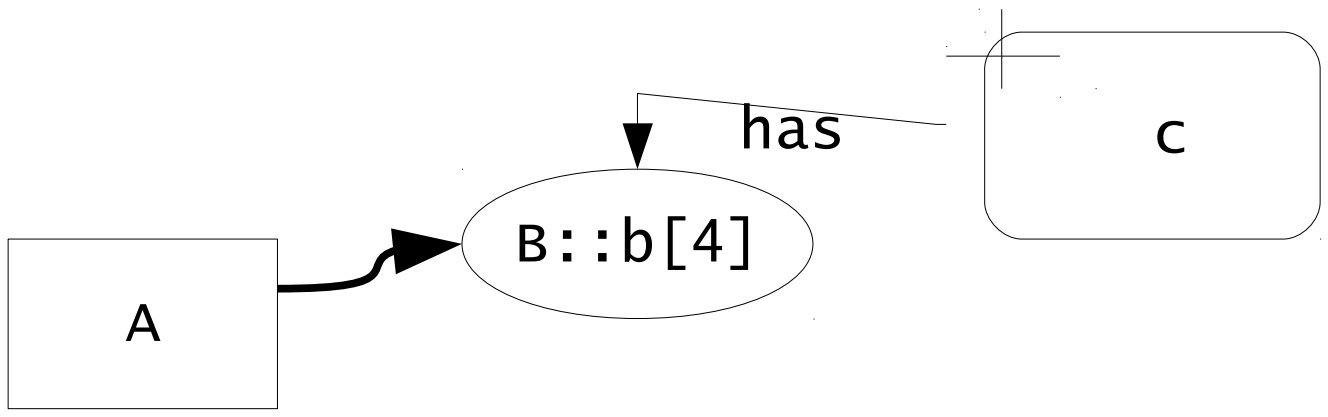

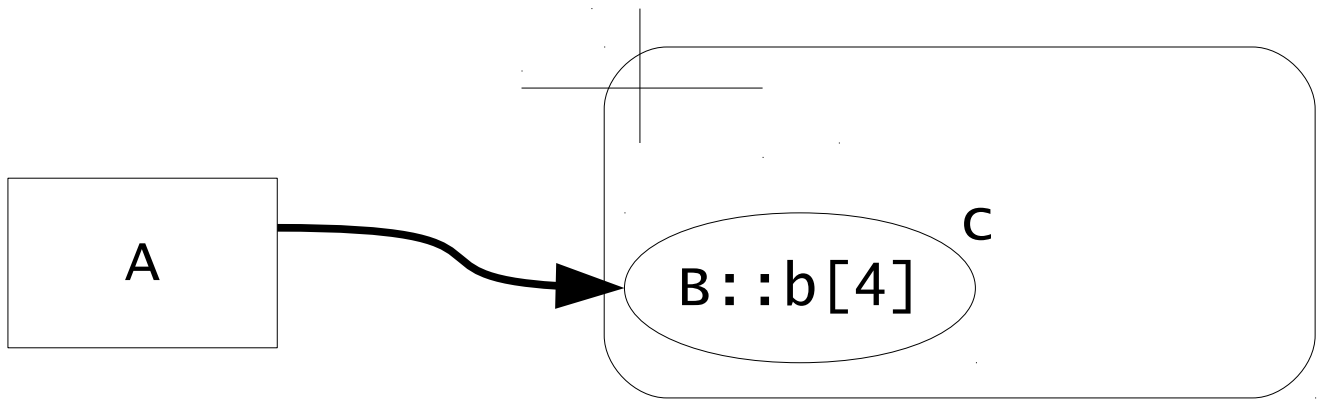

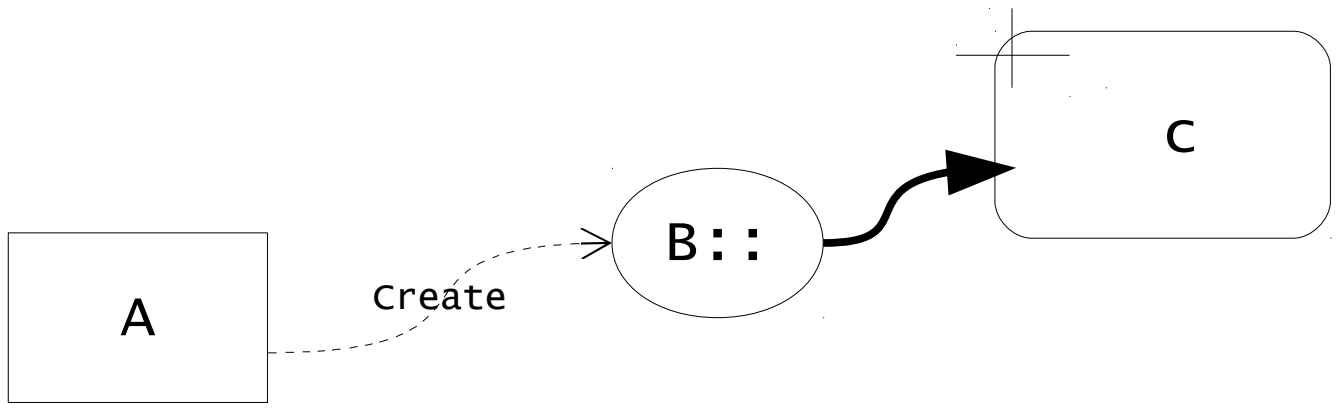

- Process A creates a record of type B in the collection "c".

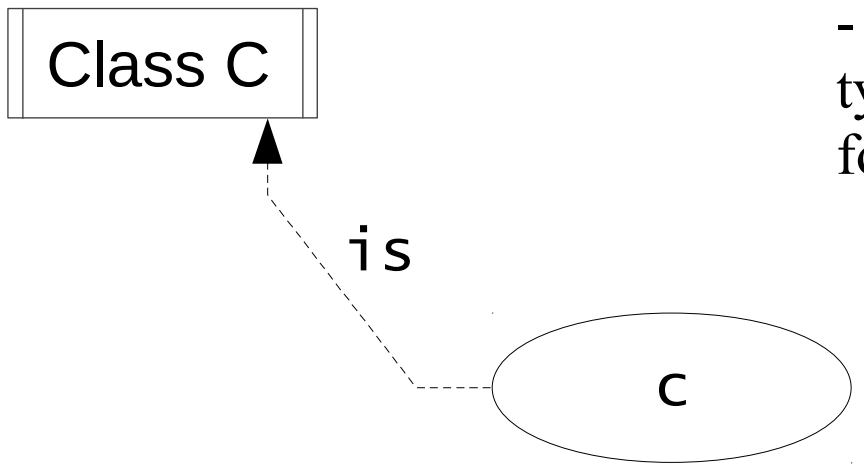

- Sometimes it is useful to show that a certain data holder is of a certain type. It is shown with an angled line representing a relation "is". It follows that "c" contains all the members of a class C.

Sometimes two shapes represent the same entity. Then, the relation “is” is in place, but it is better to be shown with an “alias”, see later.

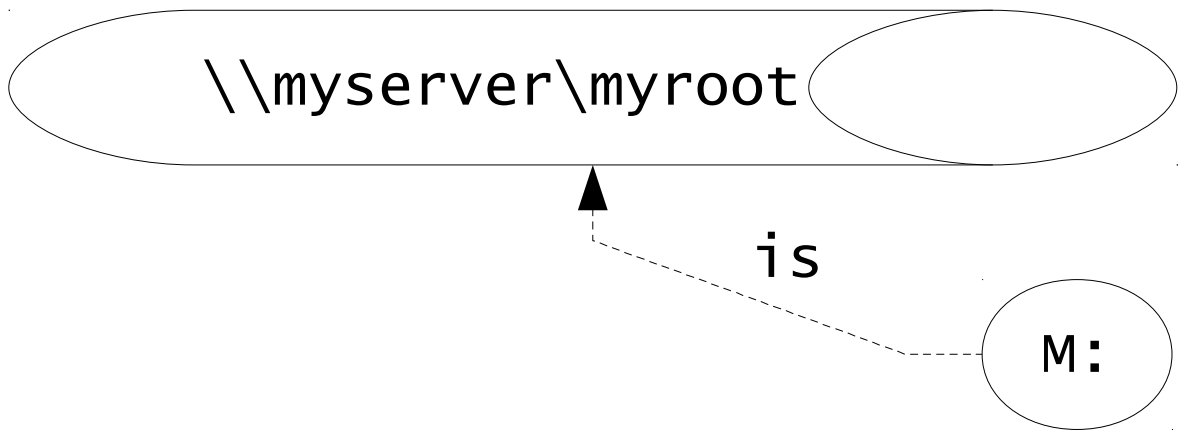

- For example, “the network path \\myserver\myroot is mapped to M:”

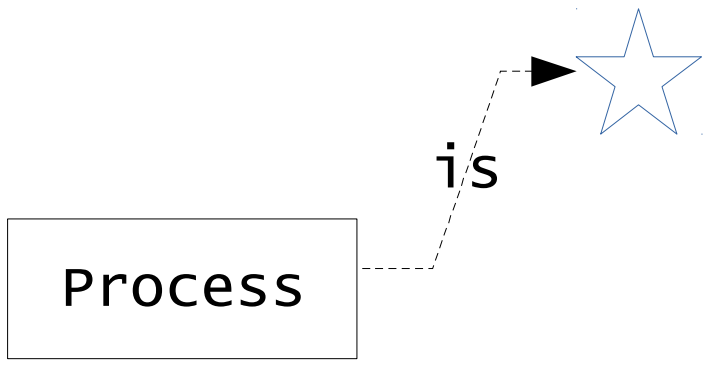

Relation “is” may be used to place various marks. For example, a star may mean “It is priority”. Other marks could be

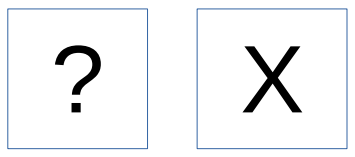

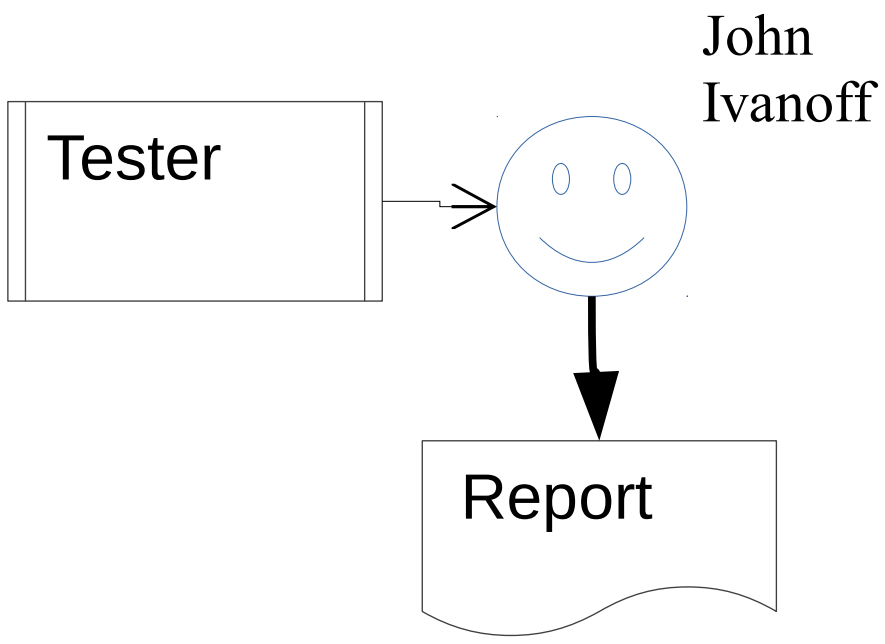

When human involvement is required, it is useful to mark it with a special symbol. For example, “John makes observations and produces a report”.

An example of a human process.

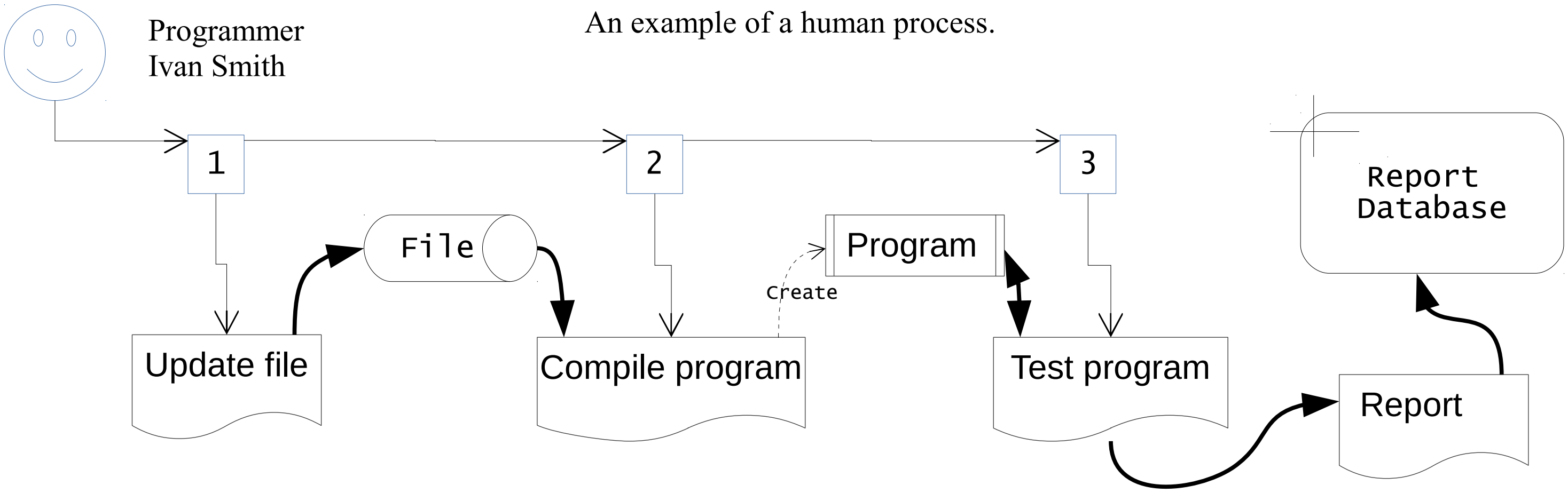

Documents can include large text with descriptions. Comments are shown as a dashed box with an angled line with a dot at the end.

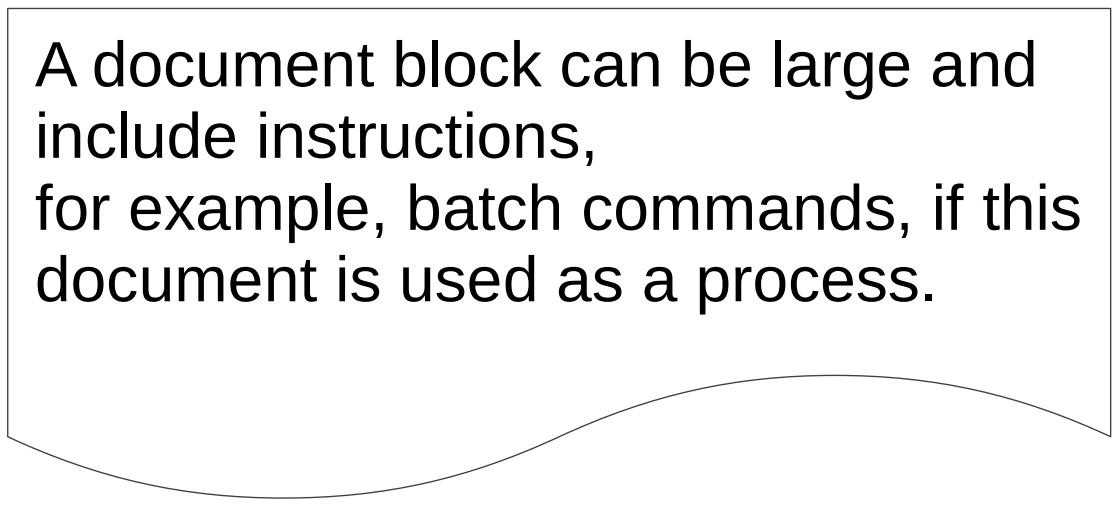

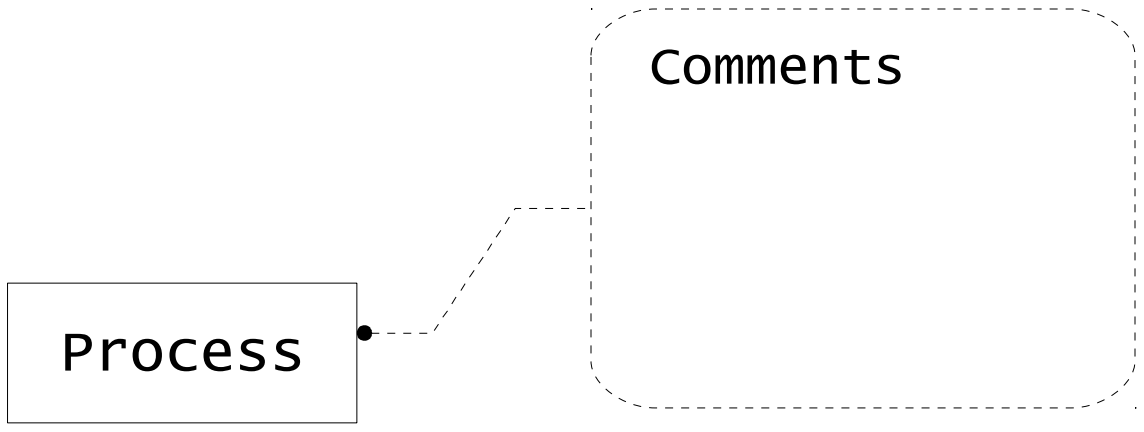

Sometimes it is useful to show that a data or a control flow can be controlled from other blocks.

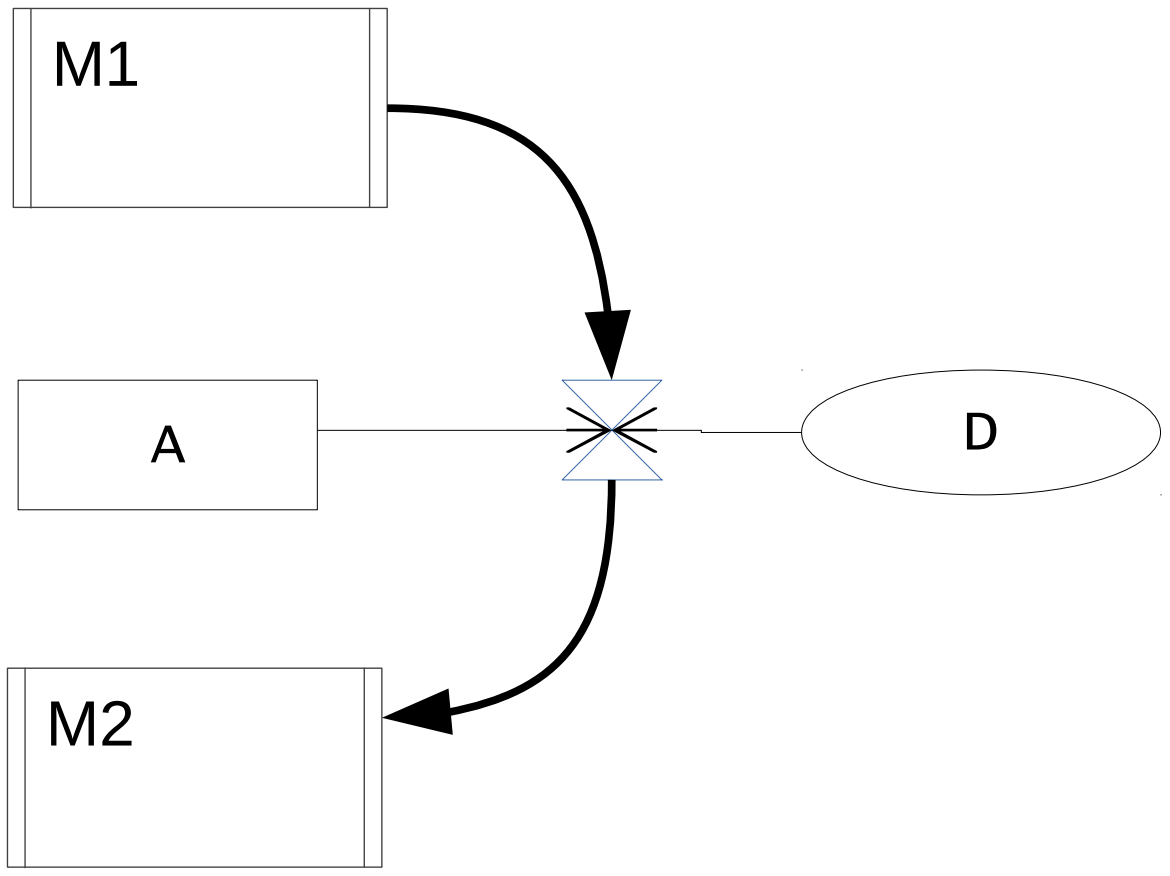

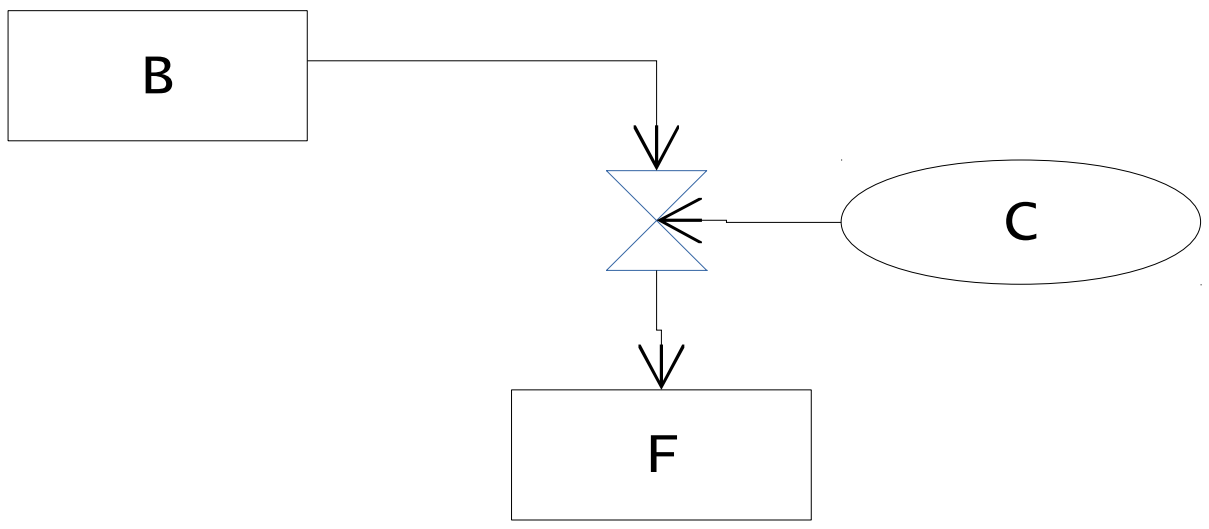

It may be needed to emphasize branches in the control flow. It is shown with a simple logical OR rectangle. This is not a process. This OR means that for some reason, the control flow may take different paths.

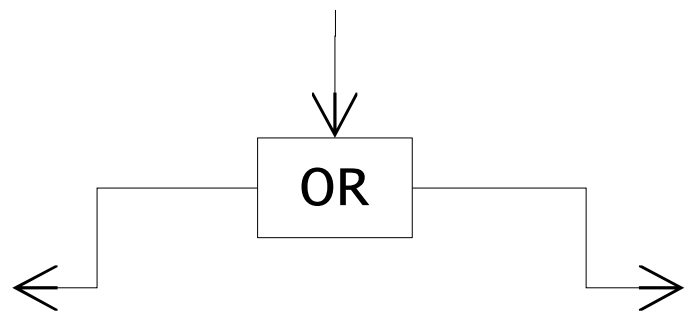

Simple logical branches are depicted with a familiar rhombic shape. This is a process.

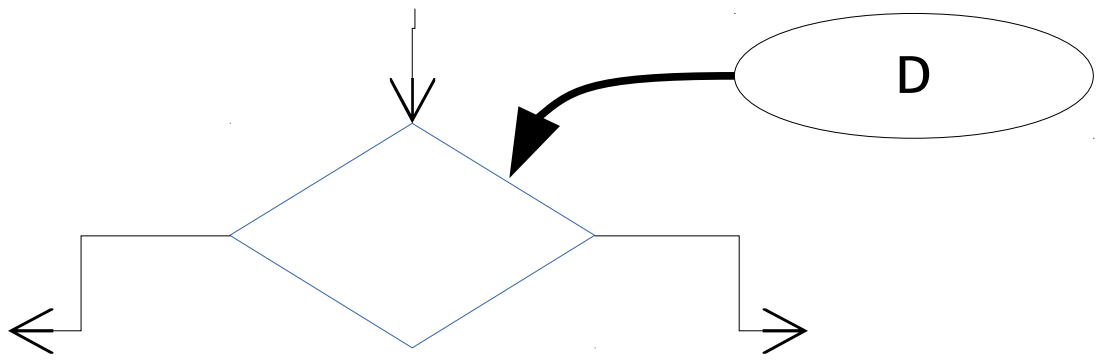

Threads are depicted with different timelines. Here is an example. Note, that in the proposed method, everything, code, data, threads are shown on the same diagram.

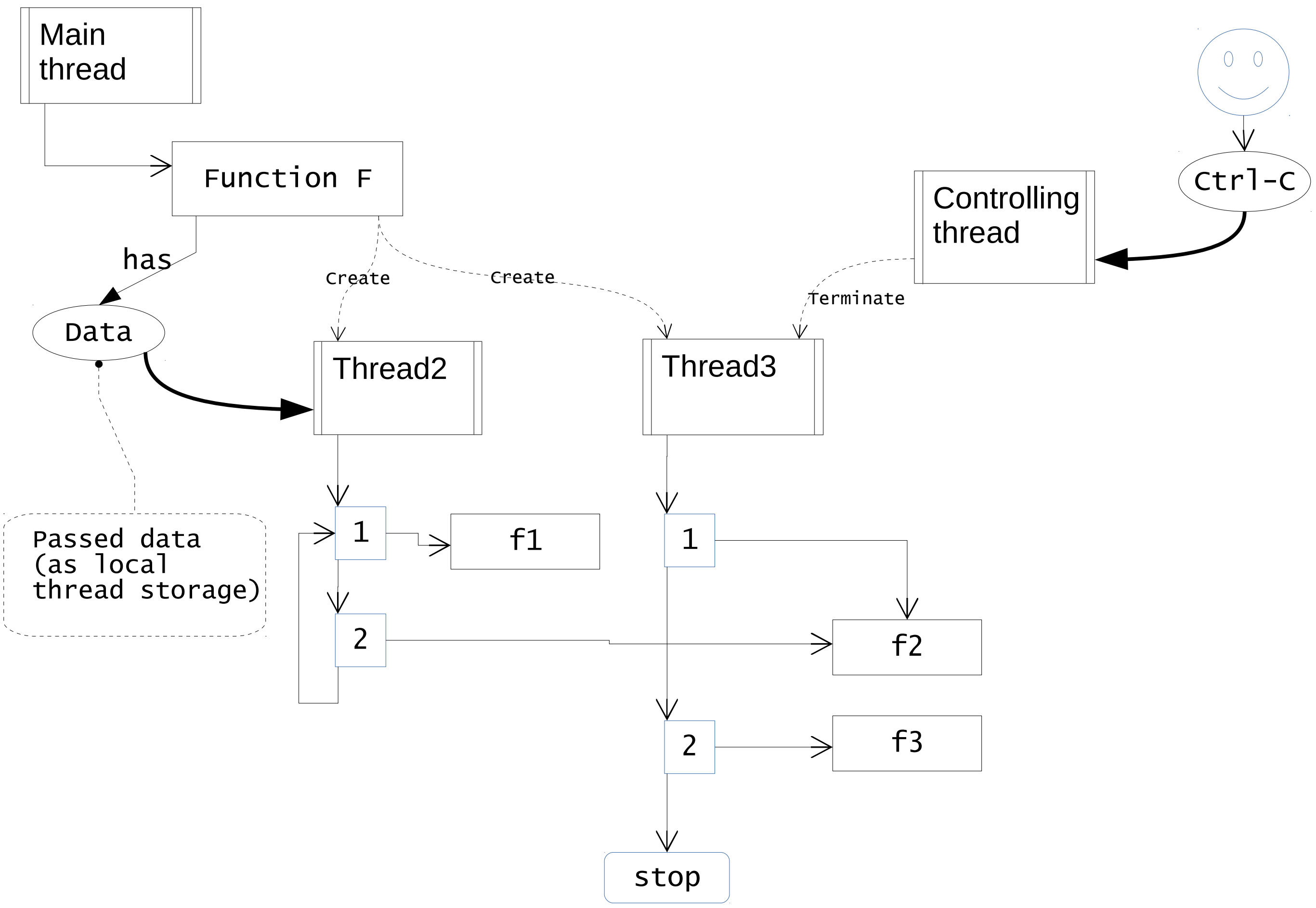

Complex processes accept and issue complex flows of information. It might be important to trace dependencies of the information flows. When the complex process is subdivided, these dependencies exhibit themselves. A real code world is far from being perfectly implemented. The diagrams below show a compact input-output diagram and a more complete diagram where dependencies are visible.

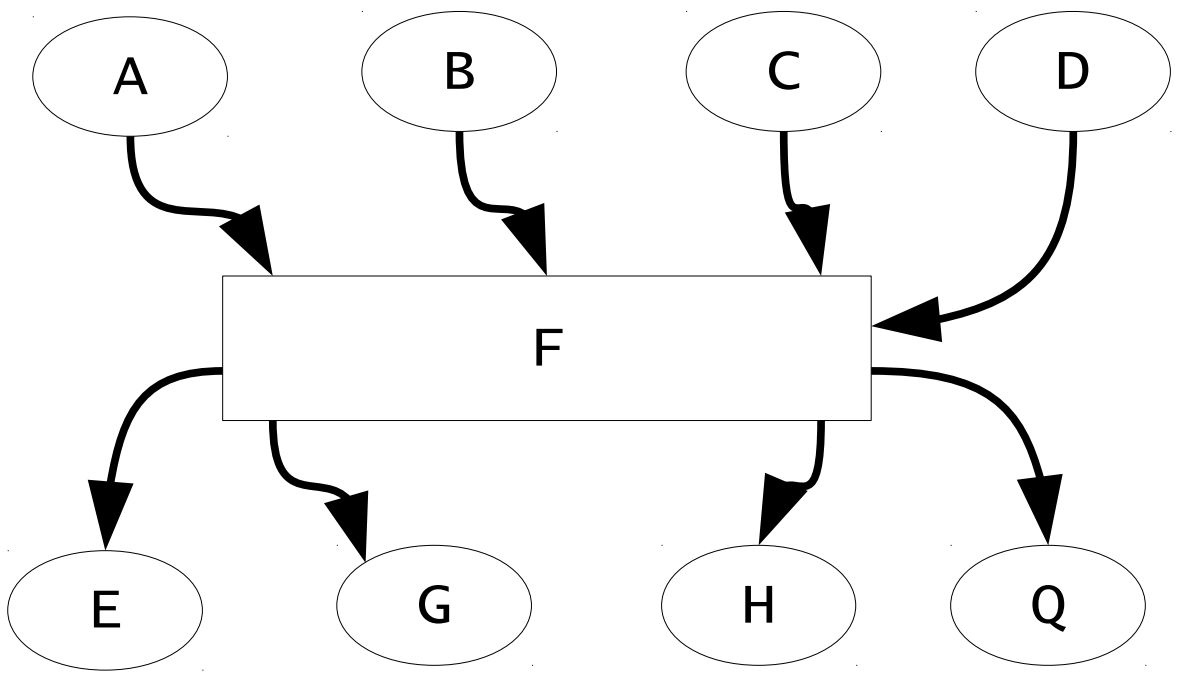

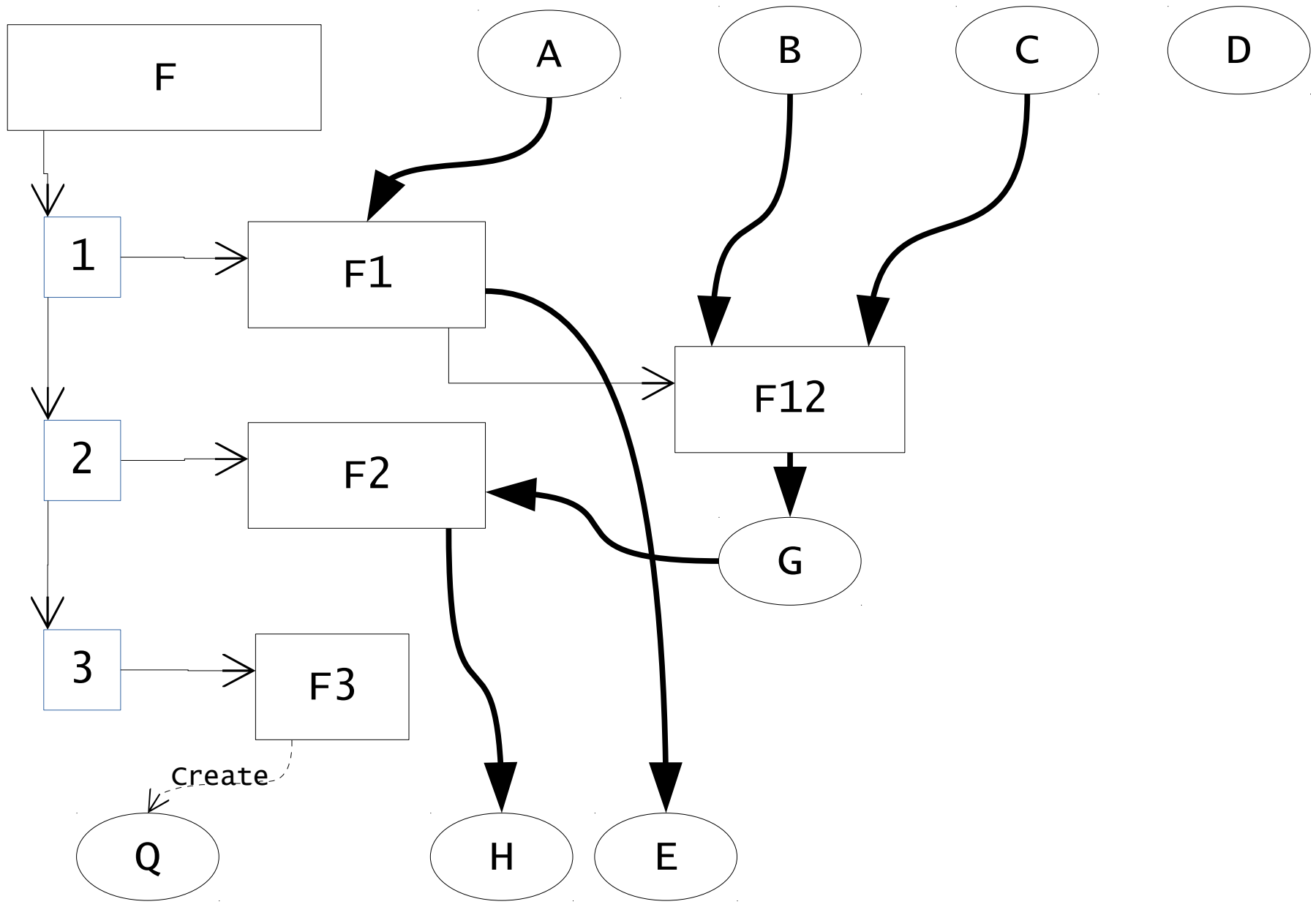

Let us look closer on the diagramming of a function call. Here is the code in C++:

```
void functionA ( arg a )
{
        arg b(a);
}
{ // Some block B
        arg a(5);
        functionA(a);
}
```

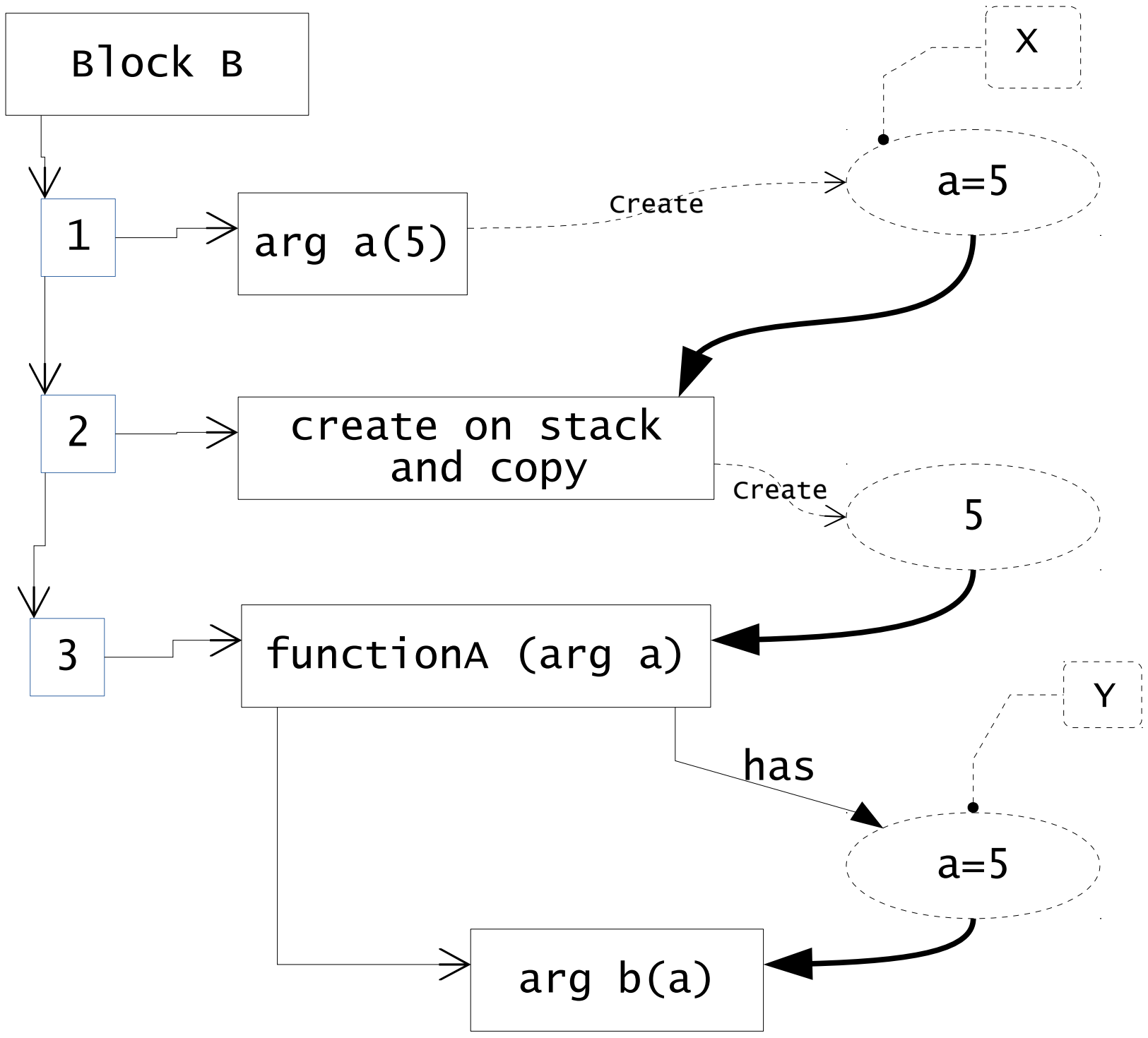

The code can be diagrammed as on the left. Note that X and Y are different memory locations, they depict different data holders.

This diagram can be simplified to avoid unnecessary duplication of data holders, since the duplications are only related to the copying of information which is the method by which a function call is usually implemented. Foremost, I want to show the flow of information rather than technical details which also can be shown if needed.

The simplified diagram is below.

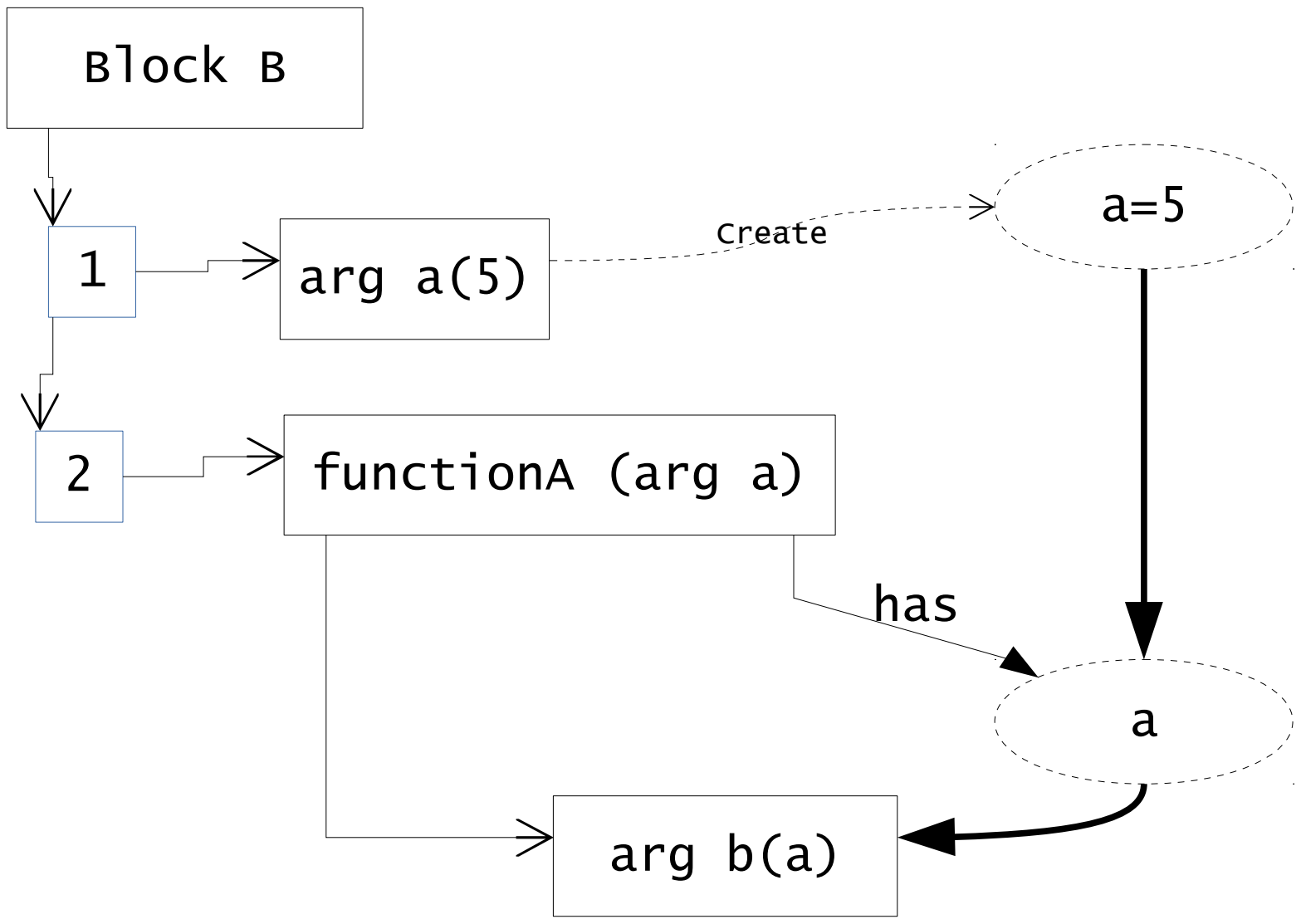

One can argue that the actual code looks much simpler than a diagram. In the example above, yes. The advantage of diagramming becomes clear when control and data flows, that are coded in different source files, are depicted. The "far" dependencies are difficult to notice in the source code, they become immediately apparent when there is "a line". Our vision is adopted to notice such things. The diagram can depict the "far" dependencies near each other and, therefore, help in understanding the code.

The diagramming helps to distinguish important and unimportant details when the code is full of them: important things can be enlarged (like with larger font or shapes) and separated. When the code is messy, diagramming helps to highlight the major parts, flows, and dead ends. When the code is long, diagramming can easily compact it into logically perceptible blocks. Diagramming can highlight, without lost of compactness, the major functionality or dependencies which are necessary to observe.

The major advantage of diagramming as I can see it, is an ability to show an "overview" of the whole code. I mean actual code. The small parts may be too small, but the overall network of parts, "densities", clustering can be seen and percepted. This gigantic graph can be "alive" changing every time a new file version is checked in. It can be alive in a more dynamic way when it animates the actual execution of the code. It should be a fabulous movie!

The step to diagramming was made long ago with all these BEGIN END and {} blocks, indents. It is called "structured programming". This is nothing but a primitive graphical trick to separate the larger system into visually acceptable parts! We live in a much more advanced age now. Yes, we can use other powerful graphical instruments to help us immerse into the code.

.

Relations “has” and “is” is often convenient to convey with Euler diagrams. Especially it is useful with function arguments, since they, as data holders, are located near the body of the function. The function call diagram above may be further simplified this way:

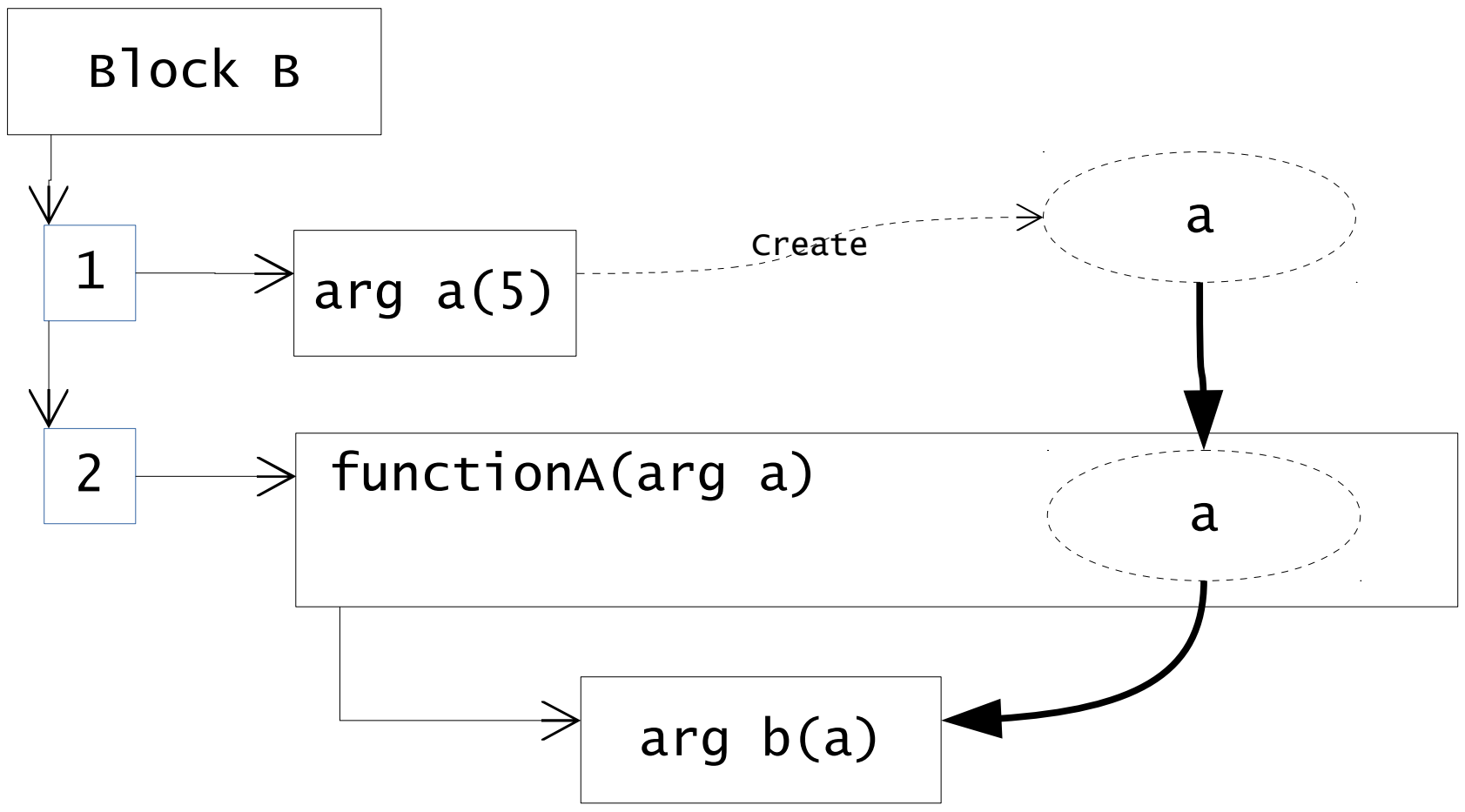

Both ways, with a “has” arrow and with the Euler diagrams, should be valid. These methods are opposite to each other methodologically. In different situations, one way is more convenient than the other.

Also note that on this diagram the process “arg b(a)” is called form within the functionA, and “a” in this case is an argument. It is clear from the drawing.

A very similar diagram is created with a member function of the arg class. Diagram systems must not rely on the language used, since all informational processes have the same underlying design.

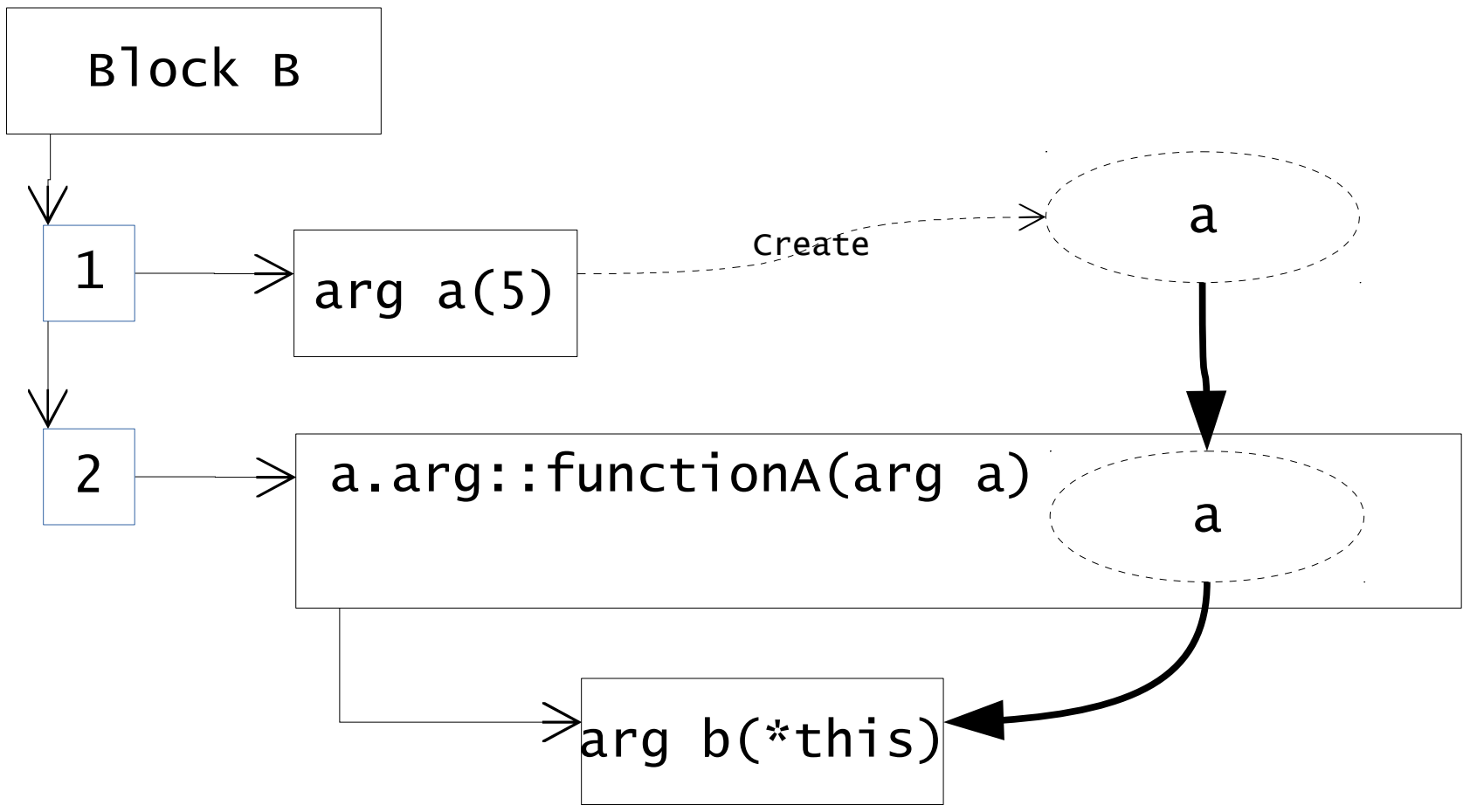

Note again, that these two “a” data holders are different. The first is an object “a”, the other is a parameter of the functionA.

Also note that on this diagram the process “arg b(*this)” is called form within the fucntionA.

C++ note. Here, arg::functionA is not necessarily a static function. arg is only a class name the function is a member of. This block can be connected with a “has” line to an arg class module block if there is a need. Here I’d like to mention that a text within a block can be any text describing the functionality of a block, including the actual code, may be several lines of it.

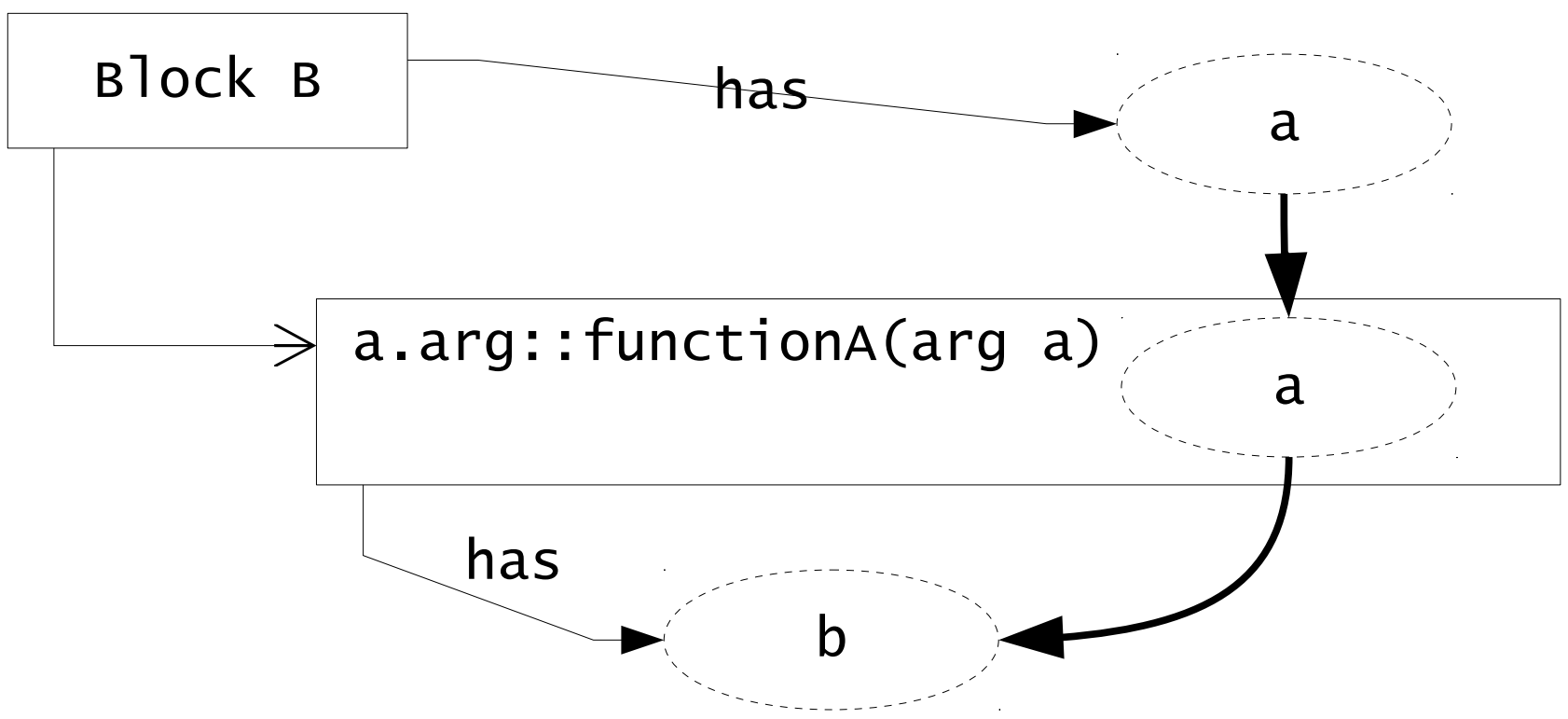

If the variable creation is irrelevant, the diagram can be simpler, like here. In this case, depicting b with a Euler diagram method would be confusing. One would need to think is it an argument or not? It is possible, though.

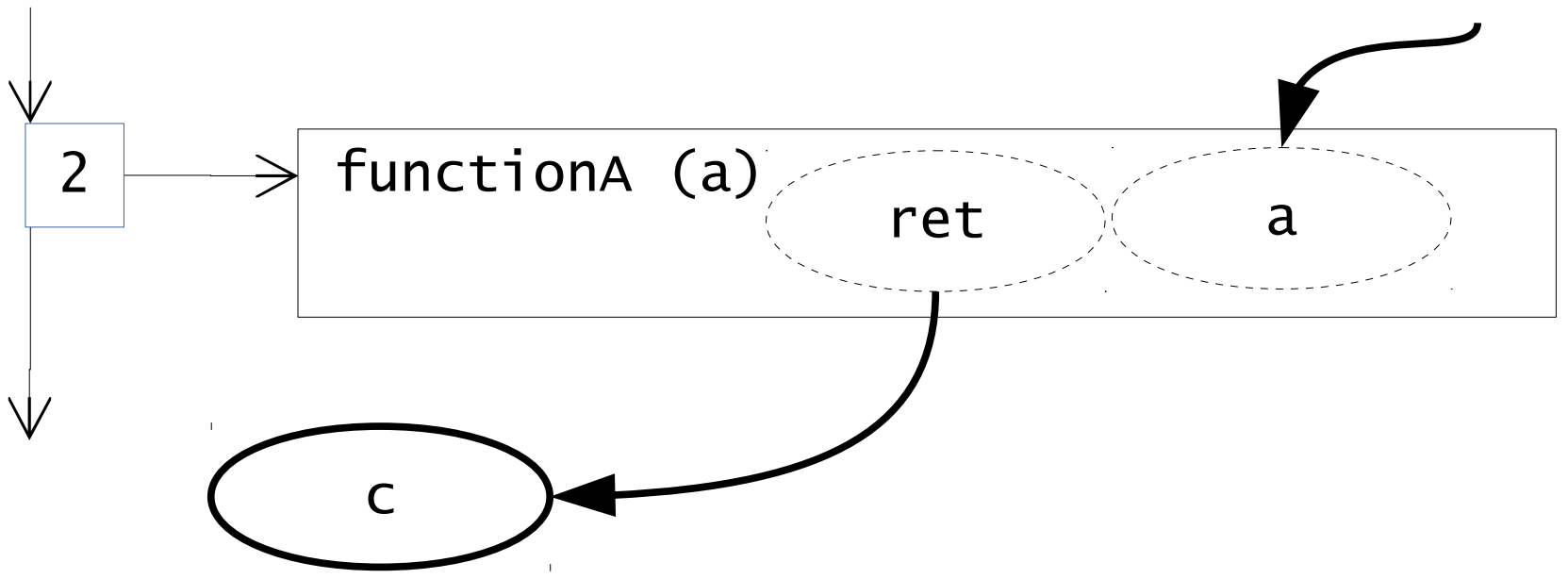

Return values as well as other output values can be easily shown with the same principle. For example, on the diagram, return value ret is saved in the “c” variable. This variable has a relation “has” with the outer block. It is not shown here, but it could be. The variable ‘c” is shown as a static variable, but likewise it can be a heap or a stack one. “ret” may have a type specifier. It will never have an input arrow.

Here is the example of a callback function. The code is

```
void func(int i)
{
        int x = i;
}
{ // calling block
        void (*fp)(int);
        fp = func;
        (*fp)(5);
}
```

The diagram representing this code could be like this. This is also an example of an assignment operator. The “func” is a static constant, since the address of any function is known.

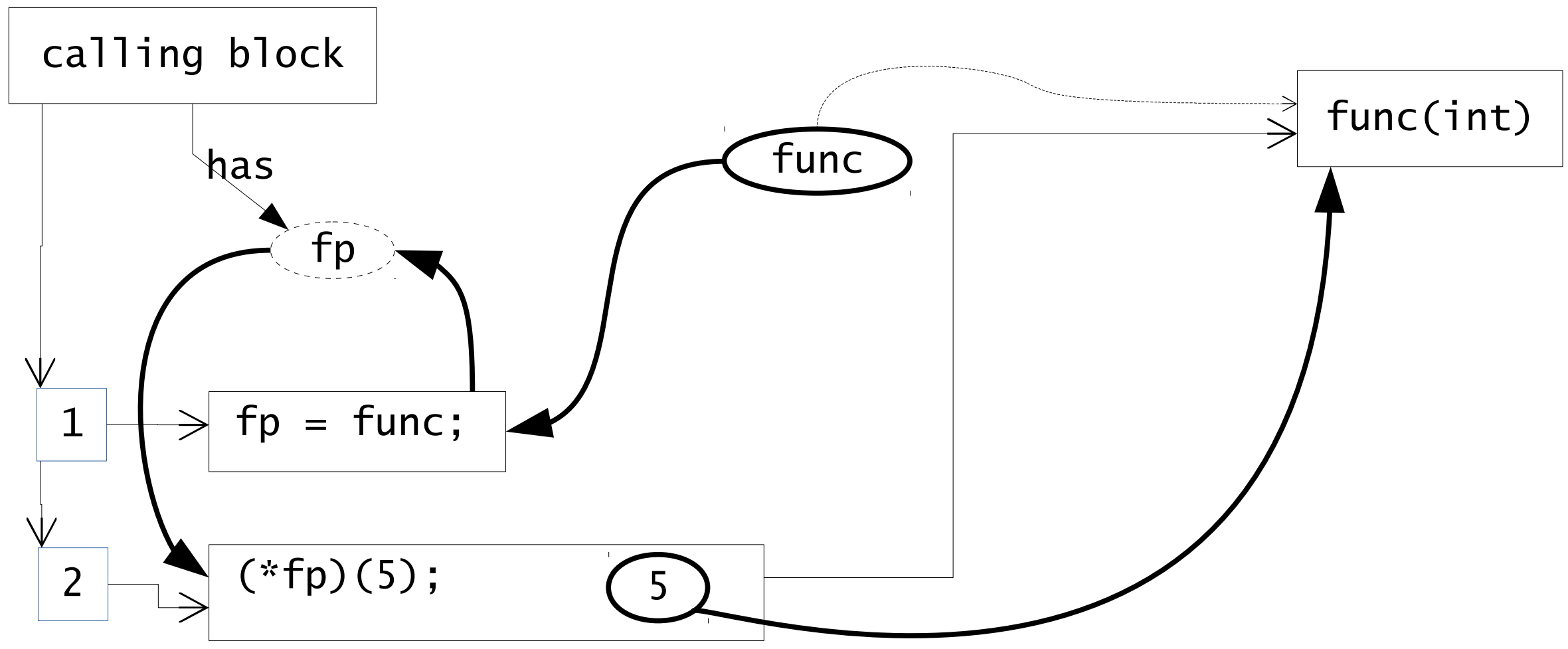

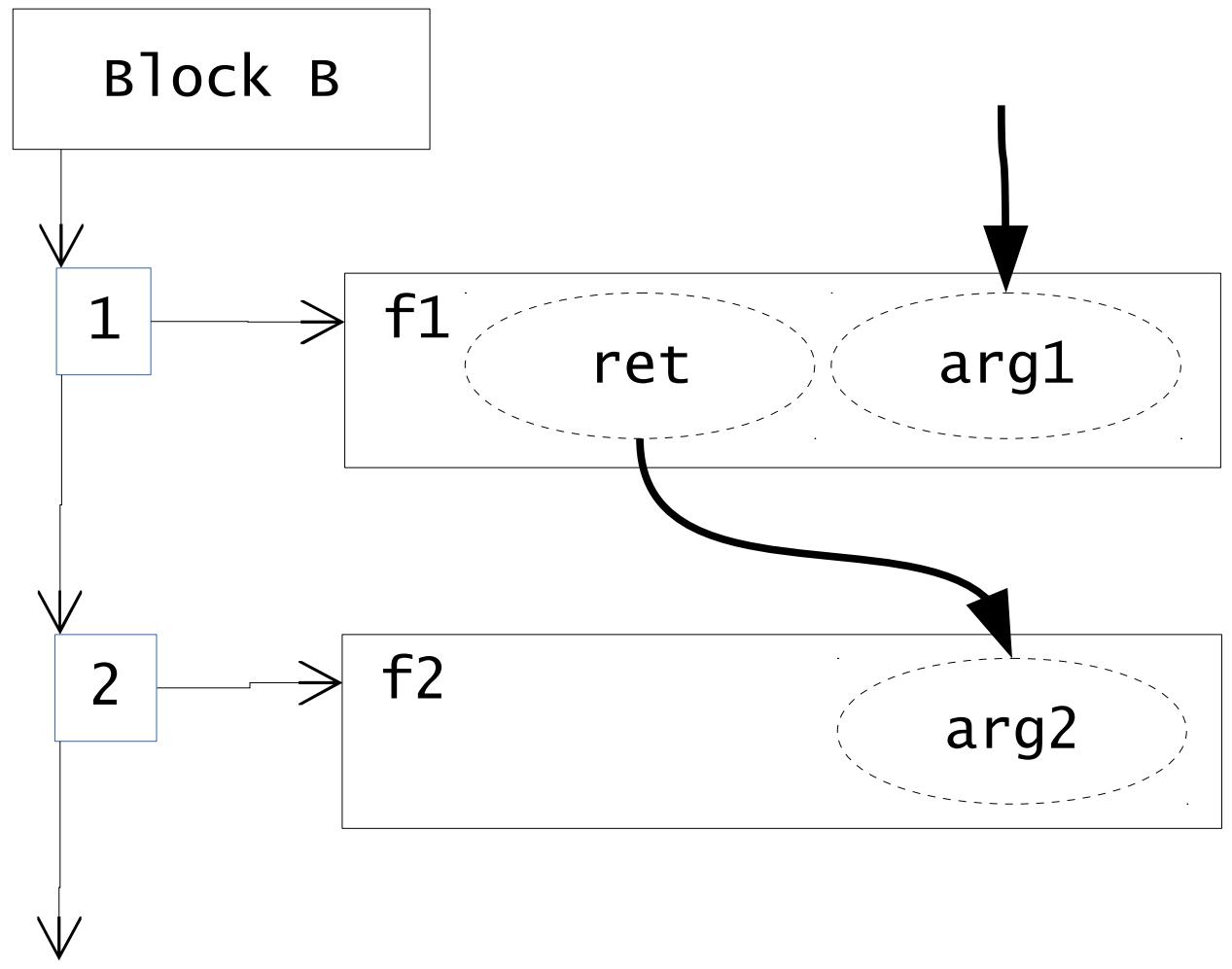

Here is an example of nested function calls
f2 ( f1 (x) );

As it was mentioned, the same function can be "called" from several timelines.

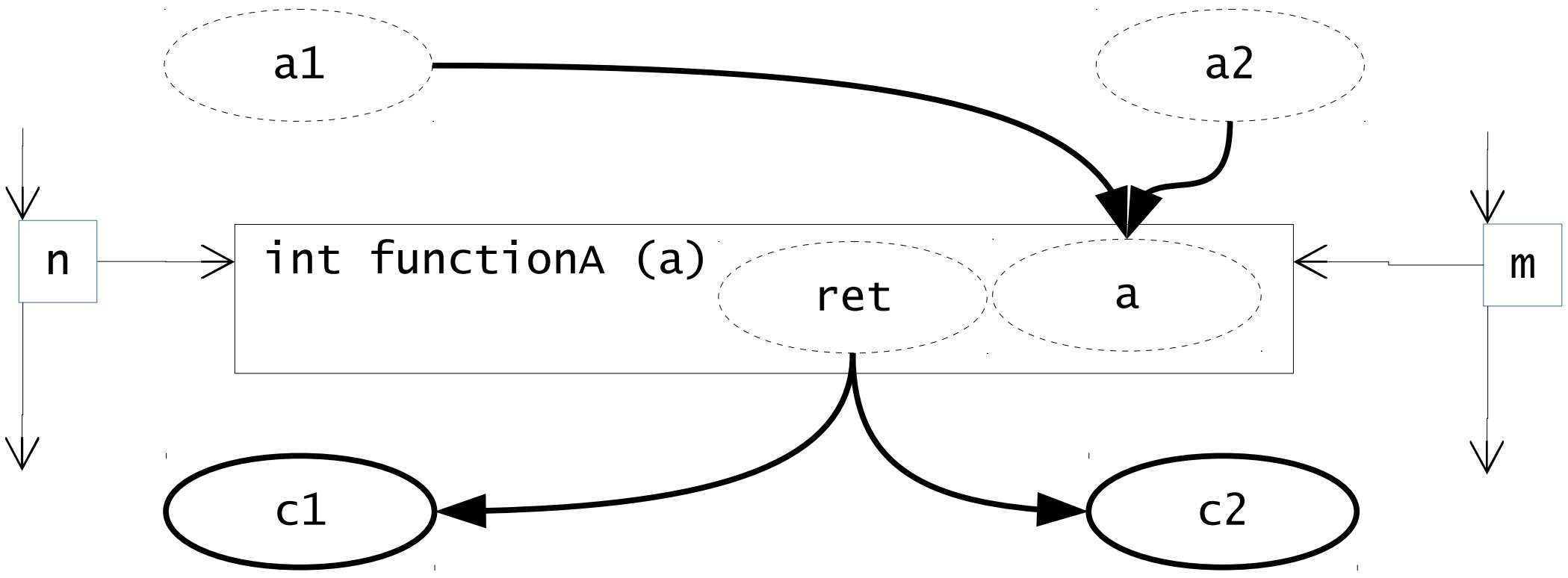

When the number of control flow lines gets higher, the diagram becomes messy. The solution for that is what I call an "alias". The above diagram with aliasing will look like one below. The complete structure of the process block is better to be presented in a designated place, while aliases may be surrounded with some internal structure required to understand the processes on each timeline. Alias relation is shown with a dot-and-dash line. If applicable, a number can show how many aliases, i.e. how many times a function is called. The same alias line can be used for data holders.

If diagramming is automatic, a complete structure may be toggled, instead of an alias, at any time.

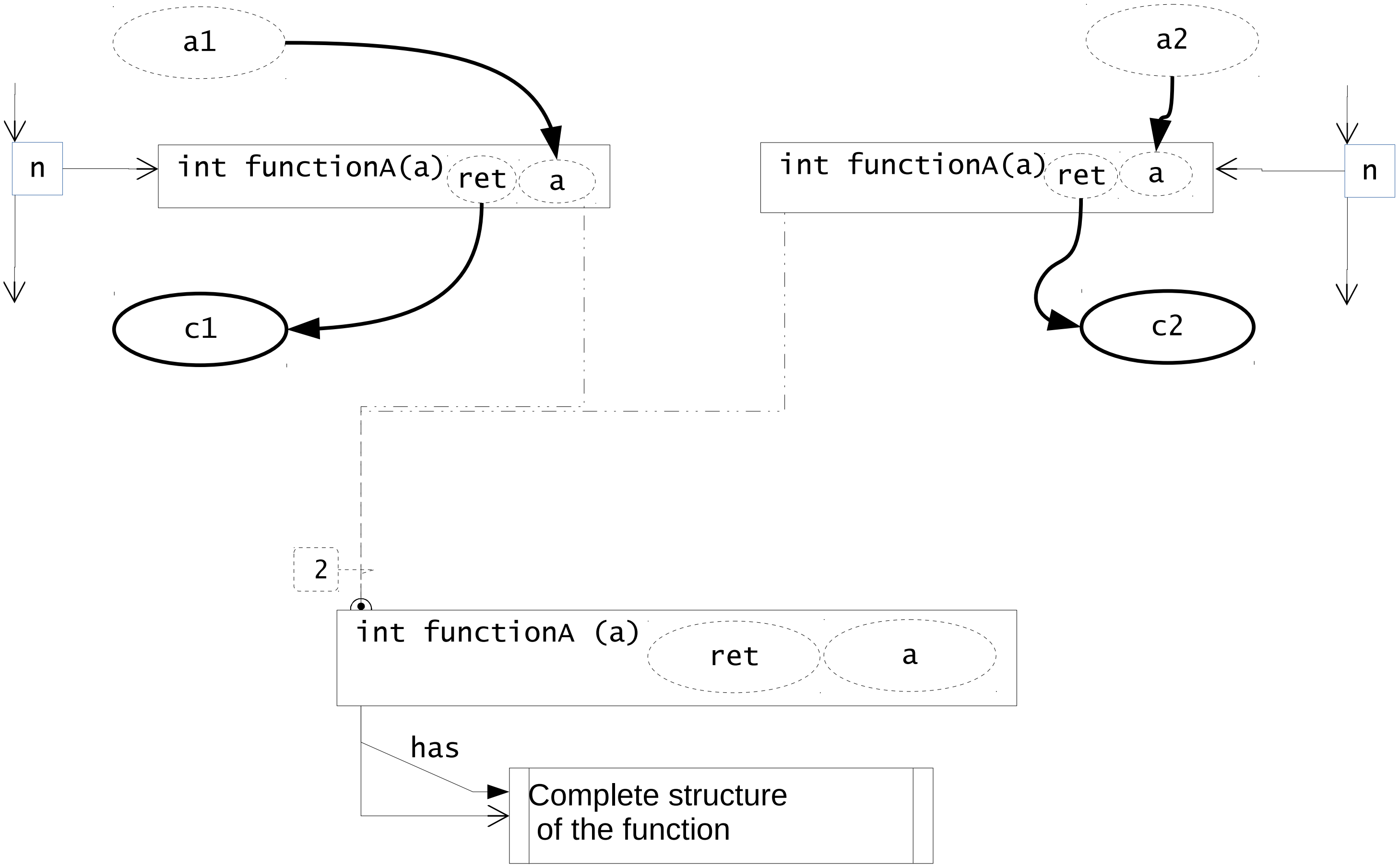

Try-Catch functionality is drawn in the way shown. The arrow from try to catch shows the control flow during exception. It is generated by implementation mechanism of the try-catch.

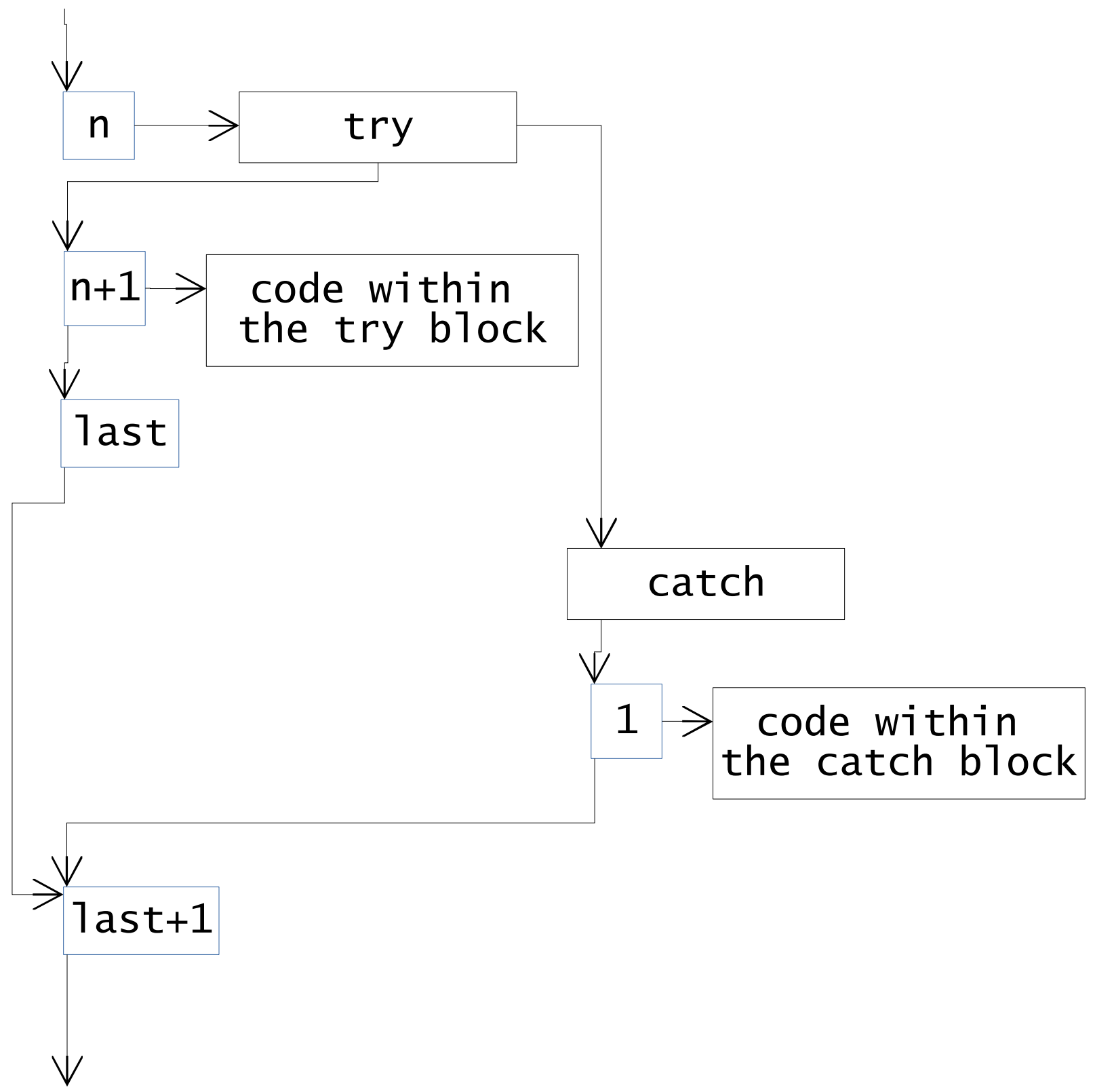

It may not always be useful to diagram a simple code, using a programming language may be clearer for a programmer, but the ability to diagram at any systematic level serves as a proof of concept that the proposed “control flow - data flow” diagramming is robust. Presenting module-level, class and function level, and operator-level code in the same uniform way is powerful.

Amendment. These are the abbreviations used in the text. This page can be used as a stencil if this document is in the original LibreOffice .odg format.

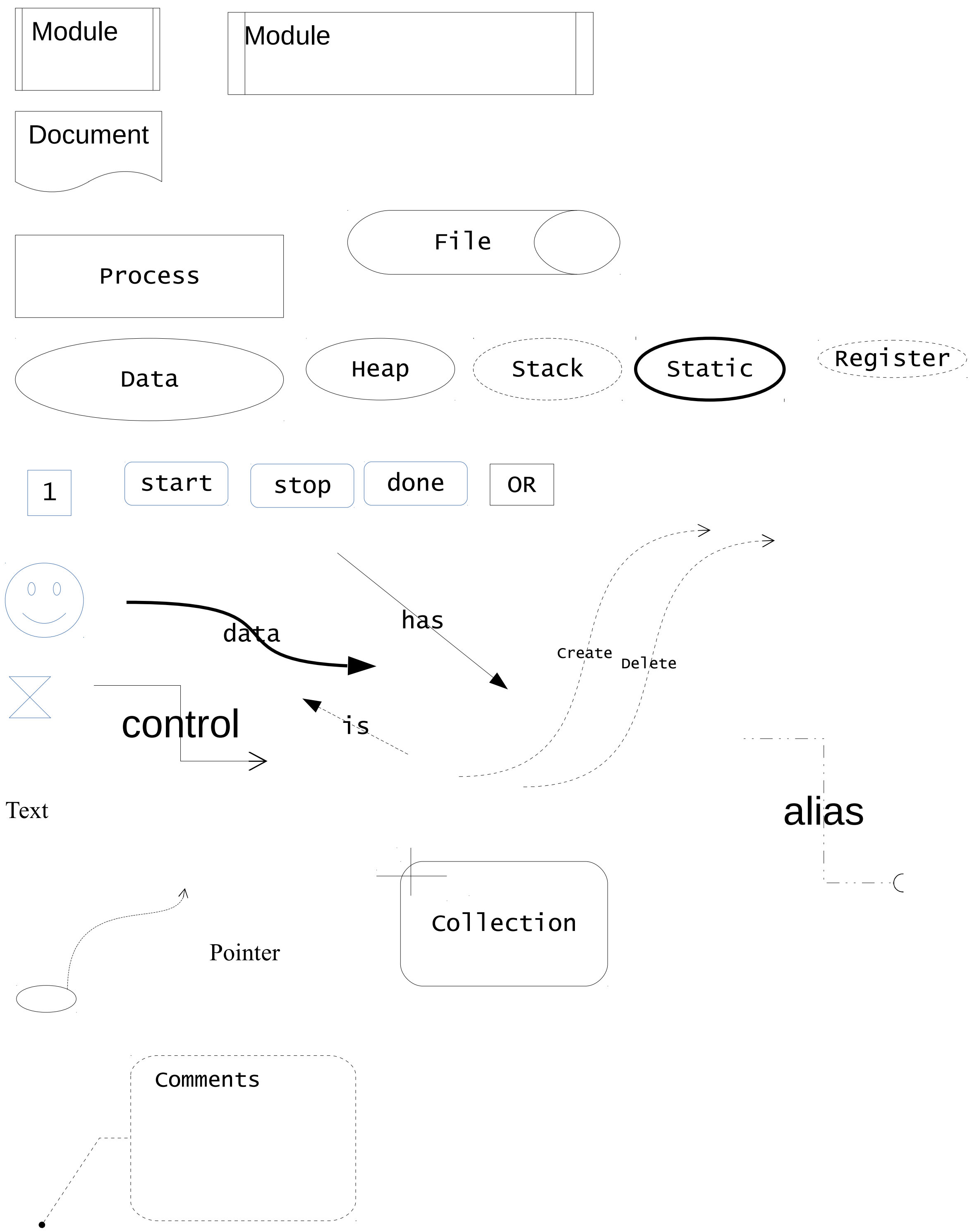

A proposed diagramming system is designed to express any information system, not only software. Here is the example of a diagram of an overview of a PTC-driven documentation process.

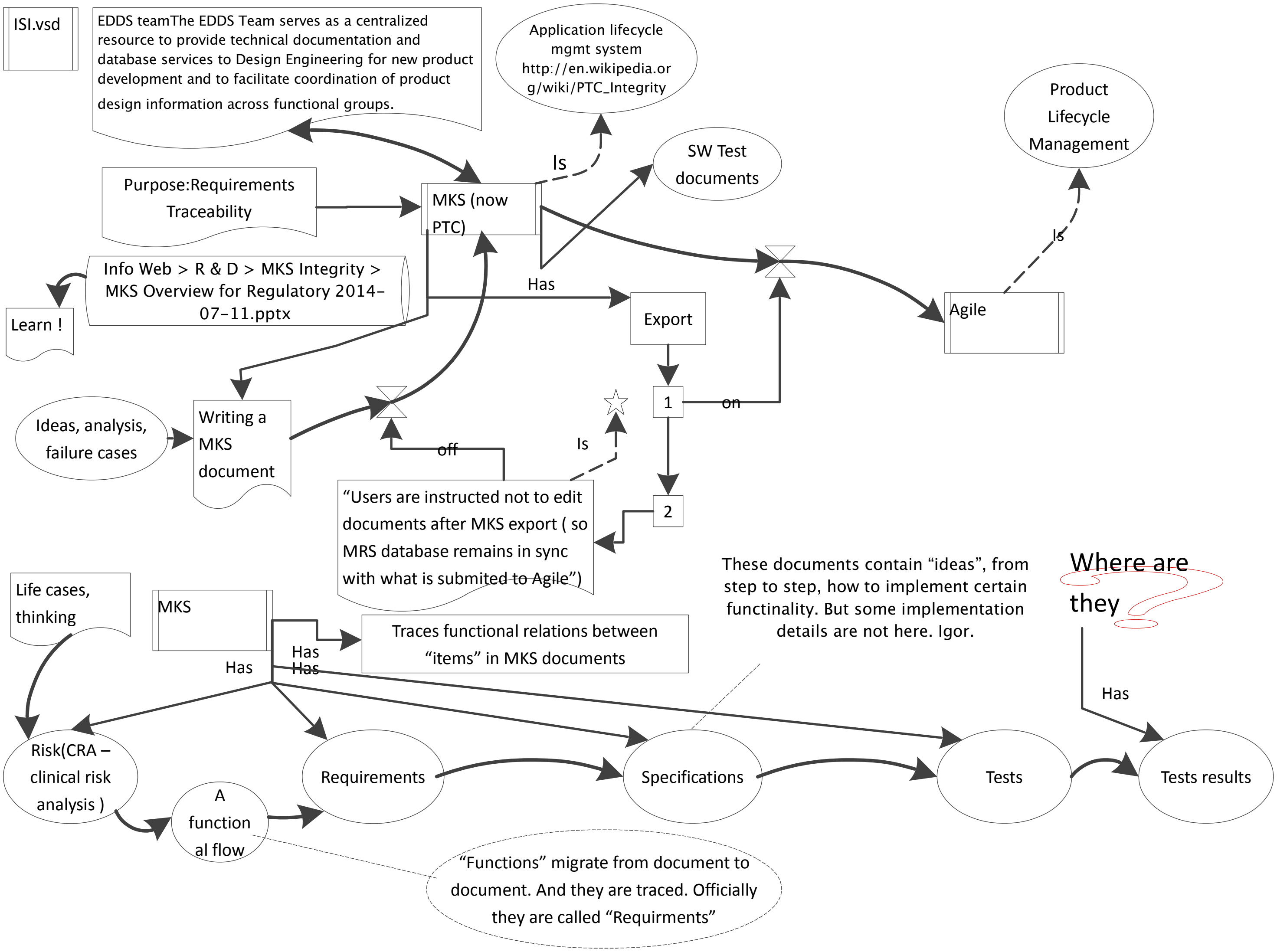

Here is an example how a production line diagram could be drawn. A product is passed along the testing stations which are shown as modules in the horizontal direction. Their sub-modules are verification programs at each station. At each station, a number "passed test step" coded in the product was updated. The main data flow on this diagram represents the flow of this information (which corresponds to the material flow of the product itself).

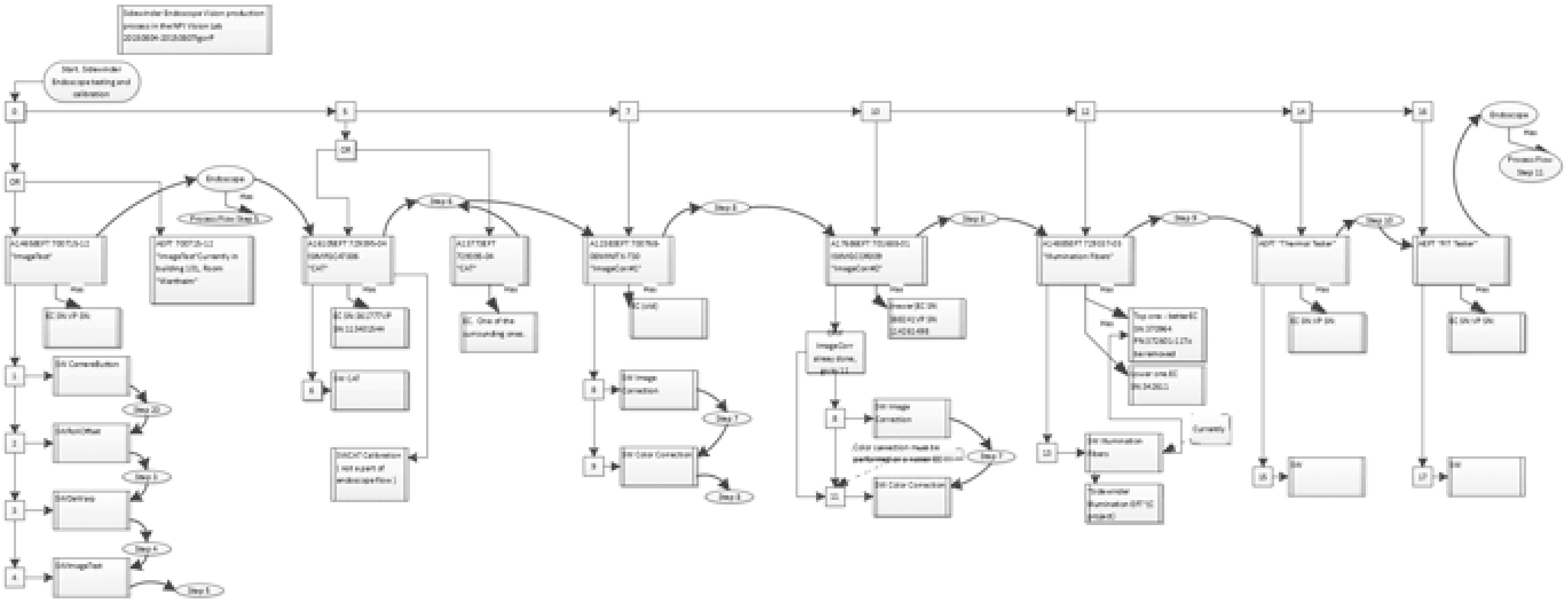

Here is a diagram of an image analysis system. It contains computers, electronic devices, ports, documents along with hardware devices like a video camera, lamps, filters, and an enclosure. In addition to control and information data flows, the diagram contains light flows and mechanical connections which are shown as thick arrows. The diagram contains all pertinent information, like serial numbers, document numbers, port addresses, and so on which are necessary for the design and maintenance of the machinery.

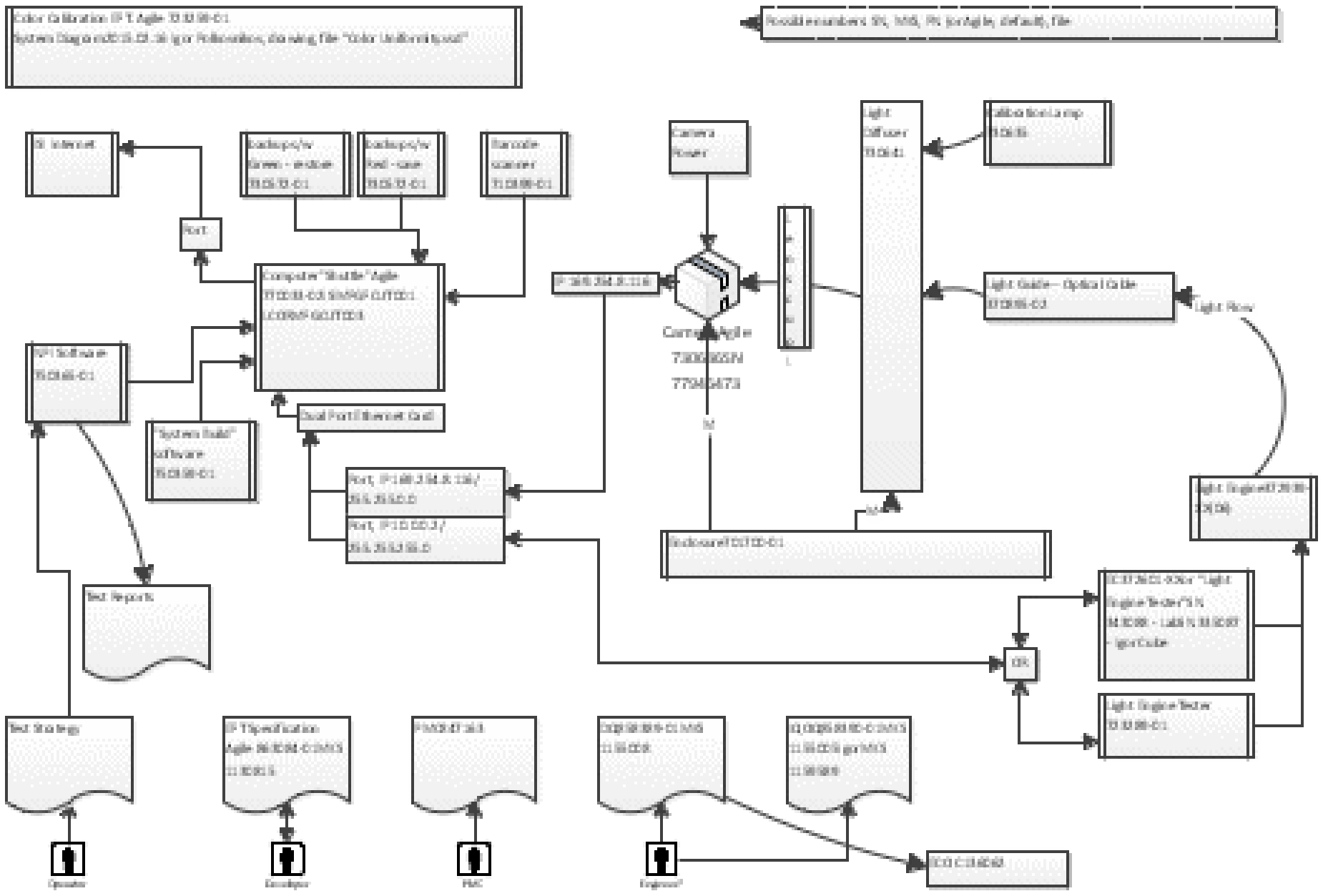

## Afterword

One of the recent attempts to address the problem of expressing actual code through a diagram of combined Control-Data flow diagrams can be found here [3]. This paper contains a useful list of the literature, so if anyone is interested in the history of the approach, they should be referred to it.

Bernhard Rumpe [4] gives an excellent overview of one current diagramming methodology applied to real-life software engineering. I'd like to note that my approach is about diagramming the actual, existing code and systems. I was not thinking about modeling nor "visual programming" and can say nothing about it, but since the approach works with the real code, I can not see much difficulty in using it as a prototype graphical language. I can recall the usage of the approach in development during evolution of a system. To make such changes in the original diagrams as amending them, re-shaping them, deleting elements, introducing new abbreviations was simple. The unified flow approach corresponds to properties of any physical system either based on flows of current, light, energy, particles, mechanical connections. But this is a topic of future works which brings us to the area of "sustained innovation"..

My work on this topic has been started at 1997. Since then I was trying to find the workable graphical representation of the actual code at any projects I was involved in. What moved me, is the knowledge of electronics. In practical electronics, a comprehensive schema of a device or a board, "electronic schematic", or a circuit diagram is 90% of success in any attempt of repairing it. In my past experience, I have fixed many different complex electronic devices without any schematic available by simply drawing their diagrams first. Coming to software engineering in 1992, I was surprised that software engineers do not use any kind similar schematics. The reason is that software is significantly more complex systematically to present it graphically. Advantage of graphical representation is obvious. It does not go away because of complexity. We have to find new ways to deal with it to the progress of humanity. Hopefully, my modest work can enlighten future developments.